\documentclass[a4paper,fleqn]{cas-dc}

\usepackage[authoryear]{natbib}
\usepackage{graphicx}
\usepackage{subfig}
\usepackage{amssymb}
\usepackage{amsmath}
\usepackage{float}
\usepackage{url}
\begin{document}
%TC:ignore
\let\WriteBookmarks\relax
\def\floatpagepagefraction{1}
\def\textpagefraction{.001}
\shorttitle{DSO-Centric Retail Market}
\shortauthors{R Haider et~al.}

\title [mode = title]{Reinventing the Utility for DERs: A Proposal for a DSO-Centric Retail Electricity Market} \tnotemark[1]

\tnotetext[1]{This work was supported in part by the Department of Energy under Award Number DE-IA0000025 for UI-ASSIST Project.}

% \tnotetext[2]{The second title footnote which is a longer text matter to fill through the whole text width and overflow into another line in the footnotes area of the first page.}

\author[1]{Rabab Haider}[% auid=000,bioid=1,
                        % role=Researcher,
                       orcid=0000-0002-5409-9769
                        ]
\cormark[1]
\ead{rhaider@mit.edu}
% \ead[url]{www.cvr.cc, cvr@sayahna.org}
% \credit{Conceptualization of this study, Methodology, Software}

\author[1]{David D'Achiardi}
\ead{davidhdp@mit.edu}
\author[1]{Venkatesh Venkataramanan}
\ead{vvenkata@mit.edu}
\author[2]{Anurag Srivastava}
\ead{anurag.k.srivastava@wsu.edu}
\author[2]{Anjan Bose}
\ead{bose@wsu.edu}

\author[1]{Anuradha M. Annaswamy}
\ead{aanna@mit.edu}

\address[1]{Massachusetts Institute of Technology, Cambridge, MA, USA}
\address[2]{Washington State University, Pullman, WA, USA}
% \address[2]{NY}
% \credit{Data curation, Writing - Original draft preparation}

% \address[2]{Sayahna Foundation, Jagathy, Trivandrum 695014, India}

% \author%
% [1,3]
% {Rishi T.}
% \cormark[2]
% \fnmark[1,3]
% \ead{rishi@stmdocs.in}
% \ead[URL]{www.stmdocs.in}

% \address[3]{STM Document Engineering Pvt Ltd., Mepukada,
%     Malayinkil, Trivandrum 695571, India}

% \cortext[cor1]{Corresponding author}
\cortext[cor1]{Principal corresponding author}
% \fntext[fn1]{This is the first author footnote. but is common to third
% \fntext[fn2]{Another author footnote, this is a very long footnote and
%   it should be a really long footnote. But this footnote is not yet
%   sufficiently long enough to make two lines of footnote text.}

% \nonumnote{This note has no numbers. In this work we demonstrate $a_b$
%   the formation Y\_1 of a new type of polariton on the interface
%   between a cuprous oxide slab and a polystyrene micro-sphere placed

\begin{abstract}
The increasing penetration of intermittent renewables, storage devices, and flexible loads is introducing operational challenges in distribution grids. The proper coordination and scheduling of these resources using a distributed approach is warranted, and can only be achieved through local retail markets employing transactive energy schemes. To this end, we propose a distribution-level retail market operated by a Distribution System Operator (DSO), which schedules DERs and determines the real-time distribution-level Locational Marginal Price (d-LPM). The retail market is built using a distributed Proximal Atomic Coordination (PAC) algorithm, which solves the optimal power flow model while accounting for network physics, rendering locationally and temporally varying d-LMPs. A numerical study of the market structure is carried out via simulations of the IEEE-123 node network using data from ISO-NE and Eversource in Massachusetts, US. The market performance is compared to existing retail practices, including demand response (DR) with no-export rules and net metering. The DSO-centric market increases DER utilization, permits continual market participation for DR, lowers electricity rates for customers, and eliminates the subsidies inherent to net metering programs. The resulting lower revenue stream for the DSO highlights the evolving business model of the modern utility, moving from commoditized markets towards performance-based ratemaking.
\end{abstract}

% \includegraphics{figs/grabs.pdf}

% \begin{highlights}
% \item Research highlights item 1
% \item Research highlights item 2
% \item Research highlights item 3
% \end{highlights}


\maketitle
%TC:endignore

\section{Introduction}
The electricity deregulation movement of the 1990s divided the vertically-integrated value chain along the power system into generation, transmission, distribution, and electricity markets. Through these regulatory changes, which include the sale of generation assets to third parties or unregulated subsidiaries, retail sales deregulation, and the revision of electricity tariffs, market competition was enabled across the value stack of the electric system. This led to more efficient pricing, free entry, free exit, and competition amongst transmission-level assets that comprise the first two value chain buckets. More recently, a similar deregulation movement facilitated competition within retail power sales, triggering the emergence and growth of competitive retail suppliers and Community Choice Aggregations (CCAs).

Within grid operations, transmission-level assets consisting of transmission lines and large-scale generators interconnected at high voltages have constituted the backbone of the power system, with a largely centralized decision and control architecture. However, this is rapidly changing with the increased penetration of distributed energy resources (DERs) into the distribution grid, which include demand response (DR), customer-sited and behind-the-meter generation such as solar photovoltaic (PV), fuel cells, electric vehicles (EVs), storage, and Combined Heat and Power (CHP) generators. A centralized paradigm may no longer be adequate with such an increased penetration. Rather, decentralized and distributed approaches are called for, as the same goals of maintaining a safe, reliable, resilient, and affordable operation have to be met by the emerging power grid. A distributed paradigm must be invoked in the economic substrate of the power grid as well, which leads to new retail market mechanisms to efficiently operate assets and support investments within a distribution system. This paper proposes an architecture for such a retail market.

% discusses the opportunity available for retail markets in building upon the initial deregulation of the power sector, to extend the transmission-level benefits into distribution systems.

Neither of the two deregulation exercises, of the vertically-integrated utility or in retail power sales, have fostered competition across distribution-level assets, and so we look towards retail markets. Discussions of DER-level markets has begun in high penetration states, such as New York, Hawaii, and California, but continues to remain in a nascent stage, with limited market-design innovation for the distribution grid \citep{NASEM_report}. The design of such markets, the operational changes, and the regulatory requirements are all open questions, for which a growing body of literature is developing \citep{Nudell2019,NRELfuture,IRENA2019innovation,MITEI_UoF}. These works look toward the establishment of a Distribution System Operator (DSO), which is charged with not only operating the distribution grid, but with overseeing the operations of a new retail market and interfacing with the wholesale market. While DSOs exist in much of Europe, they are asset-centric companies, charged primarily with a managerial role within the distribution infrastructure, and in some cases, of telecommunication, gas, and water networks. Moving forward there is an overwhelming push for DSOs to take on a service-oriented role, that includes oversight of a retail market with participation from DERs \citep{KPMG-EUoutlook}.

Distribution-level markets potentially displace agents on all revenue streams associated with the original vertically-integrated utility. In such a market, retail customers would demand and sell energy services from distributed generators with time-varying pricing (TVP). For this reason, precursors to distribution markets have taken a wide range of forms to compensate resources along the power system value chain, including TVP, retail DR, and net metering. These existing retail compensation schemes rely predominantly on static or averaged tariff rates which do not incorporate locational or temporal price differentials. As a result, these schemes fail to provide adequate compensation for DERs, whose grid services are inherently variable in location and time. TVP-based tariff structures include time-of-use (TOU) rates and Critical Peak Pricing (CPP), and have the ability to provide cost-reflective price signals and enable more efficient operation of DERs; however, retail customer enrollment in these programs has been limited. Other retail compensation schemes typically overcompensate resources at the expense of other customers, and higher future retail prices. In the absence of retail markets, wholesale electricity market (WEM) participation has also been opened up to DERs, through aggregation companies and new market models, which will continue to grow as per FERC Order 2222 \citep{ferc2222}. The WEM, however, is not designed for a large penetration of DERs \citep{MITEI_UoF}. FERC Order 2222 not withstanding, as enrollment is limited, wholesale participation models alone are not sufficient for efficient and effective DER integration, especially as DER penetration increases and a zero marginal cost system is desired. In addition, misalignment of different tariff structures between wholesale and retail programs limits the profitability for DER services \citep{SunSpecDERReport}. Regulatory and technical barriers continue to prohibit DER participation as well. Other obstacles include high costs for participation where economies of scale do not apply for DERs, misalignment in interconnection procedures between retail and wholesale programs, and rules for 24/7 participation not suited for behind-the-meter resources \citep{GundlachCAISOReview}.

In this paper we propose a retail market mechanism that aims to address these limitations, through a distribution-level market which coordinates the flexibility of DERs, leveraging the concept of \textit{transactive energy}. Defined by NIST as ``a system of economic and control mechanisms that allows the dynamic balance of supply and demand across the entire electrical infrastructure using value as a key operational parameter'' \citep{NISTtransactive}, transactive energy bridges the gap between physical power flow in the grid and market derivatives. Such a retail market has its foundation in an advanced distributed optimization algorithm, which enables local and private bidding transactions, to achieve network-level objectives. In particular, we propose a DSO-centric retail market that determines the appropriate incentives for DERs to participate in the market \citep{haiderRM}. These monetary incentives take the form of distribution-level Locational Marginal Prices (d-LMPs) to participants at the distribution primary feeders, similar to the notion of LMPs employed as pricing signals in the wholesale energy market by Independent System Operators (ISOs) at the transmission level \citep{ISO_LMP}. The d-LMPs are determined using a distributed optimization algorithm, termed Proximal Atomic Coordination (PAC) developed in \citep{RomvaryTAC,haiderRM,RomvaryThesis}, as a core component. All underlying grid physics and constraints in the distribution system are incorporated in deriving the d-LMPs. As a result, they have the potential to fully exploit the emerging flexibility of the distribution system, and reduce operational costs across the power supply chain. Technologies such as Advanced Metering Infrastructure (AMI) umbrella, ubiquitous even now, and adopted by several utilities across the US and Europe, can all be leveraged to implement the proposed retail market. In addition, an advanced communication technology that supports peer-to-peer message passing is assumed to be present to support implementation.

% in addition to the wide range of digital technologies which have already been adopted by utilities, and are yielding improvements in billing, outage response, and differentiated rate services across customer classes, will also be useful.

% , and would allow resources to meet local needs while being available to system operators and utilities to provide grid services.

% Through proper market and resource coordination between the retail and wholesale levels, existing challenges in DER integration and market participation can be alleviated. 

This paper is organized as follows. Section \ref{sec:env} provides an overview of the current US regulatory landscape, and the future environment within which retail markets are developed. Section \ref{sec:precursors} discusses the precursors to distribution markets, including existing retail and wholesale compensation schemes for DERs, and outstanding limitations in market participation. Section \ref{sec:market} describes the design of an integrated energy and ancillary services market for distribution system, provides simulation results validating market performance, and discusses the technologies required to deploy the proposed retail market design. Finally, conclusions and policy discussion are provided in Section \ref{sec:policy}. 
 
% \textcolor{blue}{Define what we consider DERs to be}
% \textcolor{blue}{\textbf{Link to transactive energy: def'n from NIST is ``a system of economic and control mechanisms that allows the dynamic balance of supply and demand across the entire electrical infrastructure using value as a key operational parameter.'' }} 
% \textcolor{blue}{benefit of better coordination of DERs as a non-wires alternative (NWA) -- allow utilities to defer investments into transmission and/or distribution infrastructure or building more line capacity, by better managing the resources on the grid and demand.}

\section{Regulatory Environment}\label{sec:env}
Integral to any plans for DER integration into market operations is the understanding of the regulatory landscape, and forms the focus of this section. We restrict our discussions to the US grid and the electricity landscape and the associated markets therein. The proposed retail market is more broadly applicable across the globe. In what follows, we provide a brief overview of the current US regulatory structure, and the future setting within which the proposed market could operate.
% in Section~\ref{sec:market} is not specific to the US, but rather an initial proposed structure, which must then be iterated upon and integrated into the regulatory landscape of the country or region -- such as the US or the EU. 

In the US, legal jurisdiction over energy and electricity interconnection, markets, and operations is divided into federal and state authority. The Federal Energy Regulatory Commission (FERC) has authority over all wholesale market operations and participation, tariff structures, generation and transmission planning, and interstate commerce, and reliability standards that are overseen by North American Electric Reliability Corporation (NERC). At the regional level, ISOs and Regional Transmission Operators (RTOs) are third-party organizations which operate the grid, oversee the wholesale market, ensure reliability and economic efficiency, and ensure non-discriminatory market participation, under the oversight of FERC and NERC\footnote{For the discussion in this paper, ISOs and RTOs are interchangeable}\textsuperscript{,}\footnote{FERC does not have regulatory oversight of the Electric Reliability Council of Texas (ERCOT), Hawaii, or Alaska, as they do not have any inter-state electricity flows, but these states must comply with NERC reliability standards \citep{NRELRegreport}}. There are seven ISOs/RTOs currently approved in the US, covering two-thirds of the US, the rest of which is operated by investor-owned utilities, cooperatives, or Federal Power Marketing Administrations. At the state level, each state has governance over retail tariffs, in-state transmission regulations, and policies for distribution system-level resources, including their interconnection policies. This taxonomy across wholesale and retail governance is particularly relevant in the discussion of DERs and their market participation.

\subsection{Relevant FERC Orders} \label{sec:ferc}
The federal regulations governing resource interconnection and market participation are laid out in the Federal Power Act of 1935 which established FERC as an oversight body, the Public Utility Regulatory Policy Act (PURPA) of 1978 and 2005 which opened market participation to non-utility generators, and subsequent Orders. As is well known, FERC has issued a series of Orders to reduce the regulatory barriers to entry for variable renewable energy generators, demand response, and storage participation in wholesale markets. FERC Orders 2006 and 792, issued in 2005 and 2013 respectively, required interconnection procedures for small generators ($<20$MW, including storage devices) and established a fast-track process to reduce the timeframe for approval \citep{NRELRegreport}. Orders 719 and 745, issued in 2008 and 2011 respectively, require all ISO/RTOs to accept bids from DR aggregators acting on behalf of retail customers, and requiring the full LMP to be paid to DR resources that are dispatched to balance supply and load \citep{GundlachCAISOReview}. In 2013, Order 784 required ancillary markets to include a pay-for-performance pricing scheme, increasing the opportunity for storage resources, and paving the way for service-based remuneration for resources \citep{NRELRegreport}. More recently Order 841, issued in 2018 and held up by the US Court of Appeals in 2020 \citep{GTMferc841}, and Order 2222, issued in 2020, address barriers for storage devices and DERs, requiring ISO/RTOs to enable participation in wholesale markets for these resources and DER aggregators, and compensate them for the services they provide. Of these, Orders 719, 745, and 841 have spurred the most regulatory activity, and Order 2222 is poised to apply additional pressure for ISOs/RTOs to revise wholesale market participation models.

\subsection{Distribution Level Regulatory Environment}
The role of DSOs has been broadly discussed in the context of existing regulatory and market structures, where DSOs typically own and operate a part of the electrical distribution network and are compensated through a rate-base model regulated by a local regulator (e.g. state public utilities commission) \citep{bos2015pricing,faruqui2012ethics,hogan2010fairness}. In many cases they also act as an intermediary between end customers and wholesale power markets by procuring electric supply. Studies such as \citep{ruester2014distribution,gerard2018coordination,BellGillReview} examine the evolution of DSOs in European countries, and report on a wide range of roles and responsibilities for the DSO, as well as coordination schemes between DSOs and Transmission System Operators (TSOs). These studies highlights the urgency with which DSOs should reform policies about access, usage, and compensation of DERs for the services they provide. Most notably, \citep{gerard2018coordination,BellGillReview} focus on emerging retail markets operated by DSOs, comparing different responsibility coordination schemes between the DSO and the TSO to appropriately operate and compensate distribution-level assets. The centralized approach is a `business-as-usual' case wherein flexibility resources are transacted within a TSO-operated market, with little to no knowledge of distribution-level constraints, and the distribution system largely continues to operate under the traditional ``Fit-and-Forget'' model. In contrast, the decentralized approach constitutes local DSO-operated markets, enabling direct purchase of flexibility resources and the aggregation of DER operation into the TSO market. The DSO-centric market structure proposed in this paper is similar to the latter approach. 

% Though the individual technologies required to develop the retail market proposed in this paper are applicable to a wide range of TSO-DSO coordination schemes and regulatory environments, the proposal of an energy and ancillary services retail market reflects a need to move beyond central markets operated by a TSO. Moreover, the proposed retail market requires that the dispatch and compensation of DERs be managed by a DSO that in turn coordinates net requirements at the transmission-distribution interface with the TSO. In the context of the current regulatory environment this would likely take place under a local regulator's supervision. It is in this regulatory environment that the DSO-centric retail electricity market design presented in this paper takes place.

% \textcolor{blue}{Discussion on FERC:
% (2005, 2013) FERC Order 2006 and 792 for interconnection procedures for small generators, establishing fast-track process and included energy storage devices in this order. 
% (2008) FERC Order 719 required all ISO/RTOs to accept bids from demand response aggregators acting on behalf of retail customers. 
% (2011) FERC Order 745 requiring the full LMP to be paid to certain demand response resources that are dispatched to balance supply and load. 
% (2013) FERC Order 784 pay-for-performance pricing was beneficial for storage, sets the tone for service-based remuneration for resources.}

\section{Precursors to Distribution Markets} \label{sec:precursors}
In the absence of distribution markets, policies and programs have been developed to compensate distribution-level resources for the services they provide to the broader grid, including direct incentives and feed-in tariffs. Many of these programs, which include DR and TVP can be viewed as precursors to distribution markets. However, these policies fall short of yielding efficient investment and operations of distribution systems, as they do not coordinate resources through bidding, dispatch, and settlement rules. In particular, these policies do not price the fine-grain locational and temporal variation in the services that DERs are capable of providing, and are therefore unable to meet network-level objectives or efficiently manage grid conditions, and struggle to maintain reliability under high DER penetration. In contrast, the proposed retail market does not have these advantages, and detailed in Section \ref{sec:market}. This section will discuss existing retail and wholesale programs, and highlight the existing barriers to market participation for DERs.
% \textcolor{blue}{and pilot projects}. 

\subsection{Time Variable Pricing}
Time variable pricing tariff structures attempt to provide pricing signals for customers to shift their consumption behaviour and more accurately recover the costs observed by the power system from final customers that are otherwise insulated from the dynamics of the electricity system.
% , and are therefore sometimes called \textit{dynamic rate structures}
Such dynamic retail rates were motivated as early as the emergence of the first failures of newly deregulated transmission-markets, such as California's blackouts and market power scandals in the early 2000's \citep{borenstein2002trouble}. In more recent years, utilities look towards TVP to reduce electricity use throughout the day, especially during times when the grid is stressed, thus reducing costs from operating fast-responding peaker plants. This translates to reductions in electricity bills for end-use consumers, and a non-wires alternative (NWA) for utilities. These rates provide more cost-reflective price signals for retail customers, and can enable more efficient use of DERs such as demand response and retail-customer-sited storage systems. 

TVP constructs include Time-of-Use (TOU) Rates, which vary the energy supply charge based on time blocks in a daily schedule, and vary seasonally; Real-Time Pricing (RTP) and Day-Ahead Hourly Pricing, which both introduce a variable electricity rate based on the underlying hourly wholesale market prices; and Critical Peak Pricing (CPP), which charges a significantly higher energy rate at times of congestion. Rate structures which hybridize these core TVP structures have also been developed, such as Variable Peak Pricing, which hybridizes TOU and RTP to have a dynamic price for on-peak blocks based on utility and market conditions; and Block-and-Index Pricing (also called Block-and-Swing Pricing), which hybridizes fixed pricing and RTP, so only a portion of the customer's load floats with market pricing, allowing them to hedge risk from price variability. 

TVP programs are designed as both opt-in and opt-out; programs that are opt-in typically have lower participation rates, while opt-out programs result in greater reduction in electricity consumption through collective buy-in power. Voluntary TOU programs exist for residential and C\&I customers, including Eversource in ISO-NE; PG\&E, SDG\&E, and SCE in CAISO, and Xcel in ERCOT. However, the majority of the mandatory TOU programs in the US are for commercial and industrial (C\&I) customers, such as PSE\&G in PJM and NYISO; Madison Gas \& Electric in Wisconsin; and City Lights in Washington. \citep{all_TVP_DR} This limits the energy and cost savings which can be achieved. Other popular TVP mechanisms include CPP which is widely used by CAISO, and is the default rate structure for most small, medium, and large businesses, including agricultural customers serviced by SCE and PG\&E \citep{all_TVP_DR}, and Day-Ahead Hourly Pricing which is mandatory for customers serviced by ConEd in NYISO \citep{coned_TVP}, and Delmarva in PJM \citep{all_TVP_DR}.

\subsection{Retail Demand Response} \label{sec:DRprograms}
% (e.g. Base Capacity Bidding in CA's investor-owned utilities, National Grid’s Connected Solutions in MA, RI \& NY, and ConEd's Voluntary Load Reduction program in IL)
Retail demand response programs compensate for reductions in electrical demand during peak usage periods, compromised grid reliability, or during high wholesale market prices. The design of DR programs varies greatly, with some programs allowing the participation of retail customers with local generators, where increased DER output reduces net load. Retail customers are remunerated through capacity payments, which are based on the committed available load reduction, and/or performance payments, which are based on the delivered demand reductions or increased power output of DERs. Some programs also have penalties for failure to reduce load \citep{DR_PJM_cpower,DR_NHPUC}. Capacity payments are typically used to incentivize customer enrollment. The structure of these DR programs vary widely, from reductions with ahead notice ranging from 10 minutes to 24 hours; load shifting to adjust usage to times with lower network load; and automated DR or direct load control programs, in which customers allow utilities to automatically reduce load during high usage periods. In the latter, control actions include adjusting programmable thermostats, cycling AC units, turning off electric water heaters, or shutting off pumps for agricultural customers \citep{all_TVP_DR}. In some programs, the location of DR resources is restricted by voltage level, to below 69kV \citep{elpaso_DR}.

\begin{figure*}[ht]\centering 
	\includegraphics[width=\linewidth]{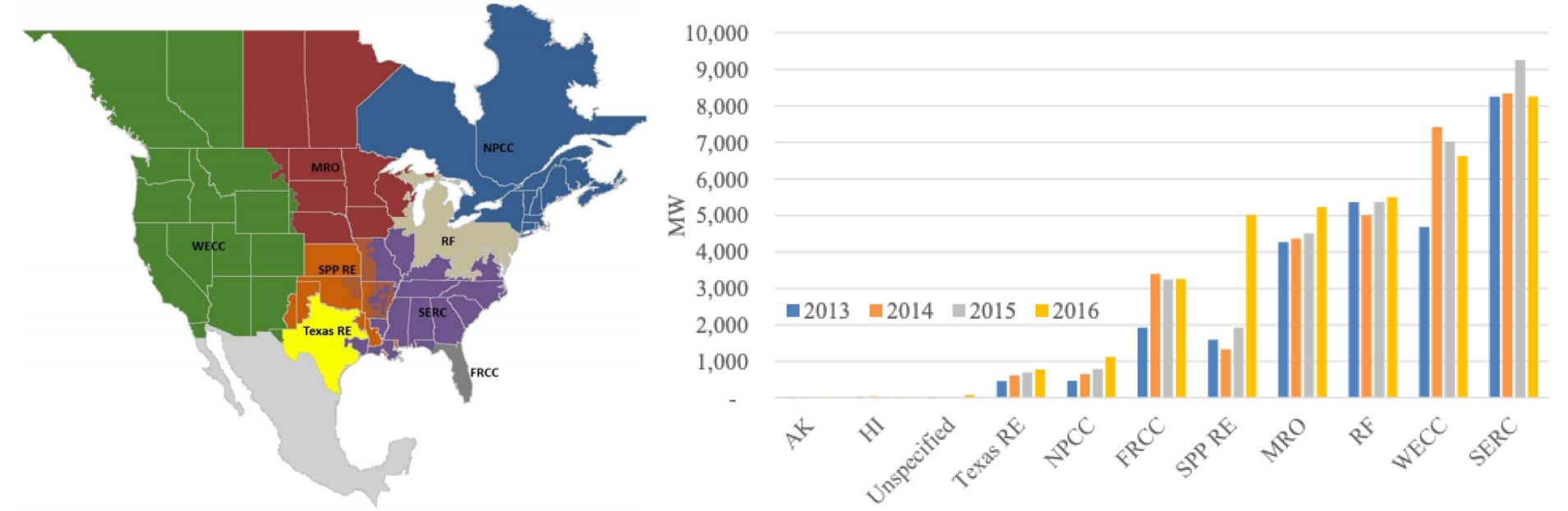}
	\caption{Retail Demand Response by NERC Region (2013-2016) \cite{fercDRReport2018}}
	\label{fig:retailDR}
\end{figure*}

Despite their diverse structure, there are several problems with these retail DR programs. First, performance payments only incentivize customers to change their behavior during performance periods or call windows (e.g. 2-6pm non-holiday weekdays June 1-Sept 30). Many programs also cap the number of times a resource can be called (e.g. 10 times in the summer months of May-August). As a result, they do not translate into continuous compensation mechanisms for DER operation across the year, with studies showing little difference in consumption outside of call windows for consumers participating in DR programs \citep{anaheimCPP}. Second, baselining methodologies are used to determine the expected consumption, and to calculate demand reductions. Unfortunately, these schemes frequently over- and under- compensate customers as they attempt to determine a counterfactual with no load reduction \citep{haasDR2017}. This method, of predicting the baseline using historical data, is not robust to changing customer behaviour or environmental events, such as weather-specific behavior events in summers and winters when thermally-dependent loads can be larger than previously observed. Further, it is prone to customer gaming, as DR participants can increase consumption outside of call windows, especially during baseline setting times, to be compensated for reducing this higher load, as done by the Baltimore Orioles baseball stadium in 2013 \citep{haasDR2014,baltimoreBaseball}. Third, incentive payment rates [\$/kW reduction] are highly variable between programs. Customers are typically remunerated at fixed prices set in contracts with the host utility, or variable prices based on the tier of capacity commitment and time duration. As such, they are largely static rates which overcompensate resources \citep{anaheimCPP}. These payments reflect utility peak demand costs, equipment upgrades, and emergency conditions, and are typically not reflective of the actual service provided by the DERs. Finally, and most importantly, retail DR does not provide adequate incentives for desired behaviours to meet state-wide RPS or energy goals. The overcompensation of DR undermines long-run investments in energy efficiency \citep{haasDR2014}, as it lowers baseline consumption and total kW curtailable load. These DR programs have not seen any clear positive growth, as apparent in Fig.~\ref{fig:retailDR} which plots retail DR capacities by NERC region between 2013 and 2016. It is necessary to note, however, that the inefficiencies highlighted above lie within the design of the retail programs, and not within the resource itself. The flexibility and distributed nature of DR poises this resource as a key component of the future electricity grid, as displayed by the demand reduction and flexibility offered to the California grid during rolling blackout conditions from extreme weather events in 2020. These success came with a shift away from traditional DR composed of large C\&I, and towards aggregations of devices and appliances such as thermostats, batteries, and home car chargers, which are dispatched by the wholesale market \citep{DRnews_CAISO}. This newer model clearly reflects the potential for DR resources in retail markets.

% More advanced DR mechanisms in ISO-NE,PJM,NYISO:

% Delmarva, MD.Peak Management Rider (Rider PM) program, gives per-kW credit for reaching pre-designated firm servicelevel or guaranteed load drop.  Remuneration based on PJM wholesale price + max days curtailment +max hours curtailment.  Non-compliance penalties apply

%  Voluntary Interruption:(1) Eversource, NH. Voluntary Interruption Program (VIP) to curtail max(10\%of load, 100kW) during summer/winter (or on-site gen), paid some \% of the hourly real-time zonal price(RTZP) and (if applicable) excess interruption credit of 60\% RTZP, with penalties for failure to comply.Cannot participate in VIP and ISO-NE DR programs

\subsection{Behind-the-meter Compensation Schemes} \label{sec:retailcomp}
% In addition to integrating behind-the-meter and small-scale DERs through load reductions in DR programs, other retail compensation mechanisms have been developed. 
In the absence of retail markets in the distribution grid, several compensation schemes have been proposed for behind-the-meter resources including generators (both conventional and renewable) and storage devices, based on passive metering and load management by end-use consumers. In regions where compensation schemes are not present, local generation resources simply reduce the customer load, and excess generation is curtailed. Compensation schemes allow exporting to the grid, for which the customer is compensated, through either net energy metering (NEM) or net energy billing (NEB) \citep{NEM_NEB}. In NEM, the excess generation (which is exported to the distribution grid) is subtracted from the imported energy (delivered by the distribution grid). This net energy is charged to the customer at the retail rate if net imports exceed exports, or purchased from the customer if exports exceed imports at a purchase rate. In NEB, all imported energy is charged at the retail rate, and all exported energy is purchased by the utility at the purchase rate. These amounts are netted and charged or credited on the customer's final bill. 

\begin{figure}[ht]\centering 
	\includegraphics[trim={0 1cm 0 2.7cm},clip,scale=0.3]{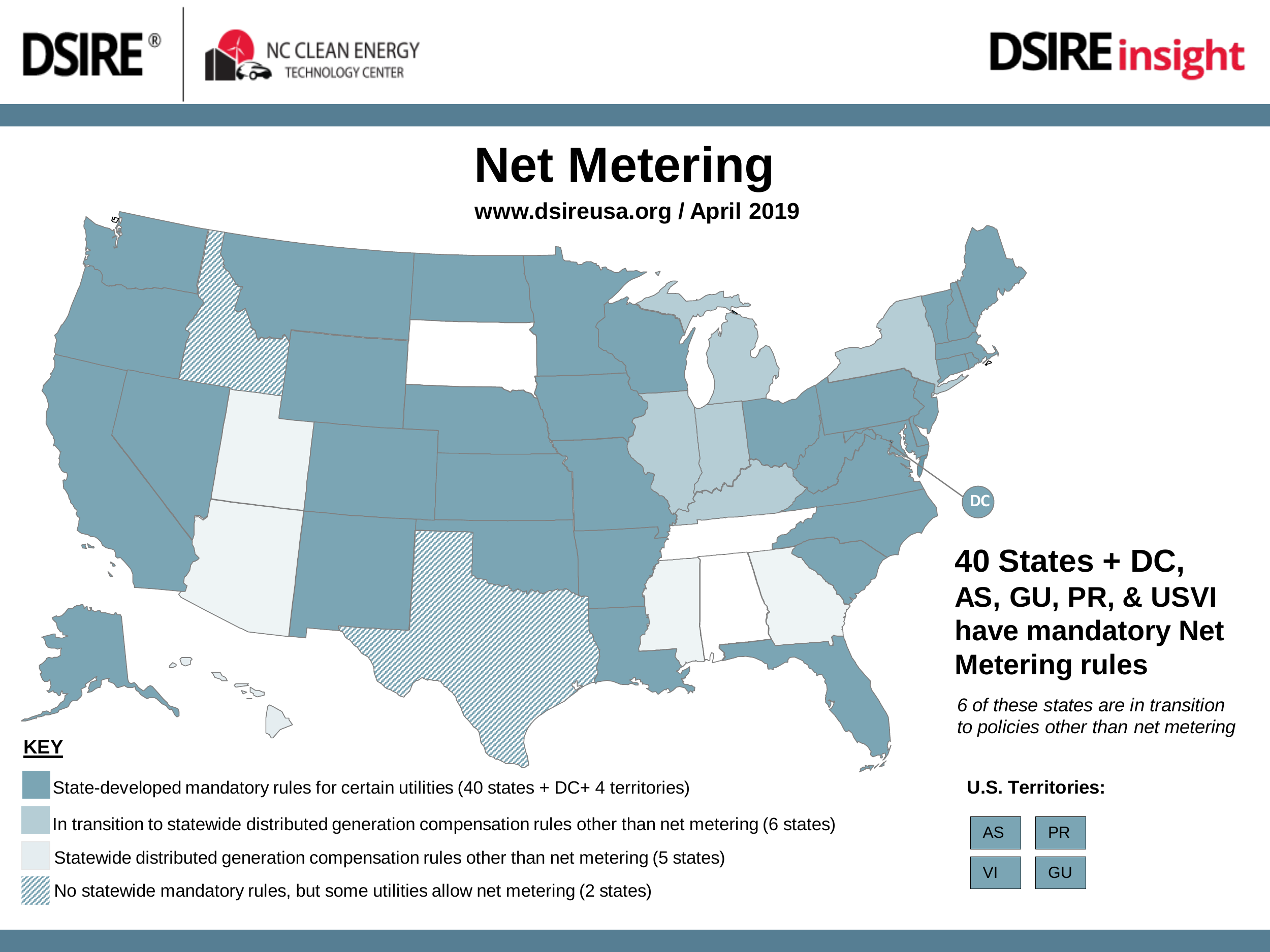}
	\caption{Net Energy Metering Policy By State (April 2019) \cite{dsireNEM2019}}
	\label{fig:nem}
\end{figure}

Many state regulators have implemented such NEM and NEB schemes to compensate DER exports. These policies range from mandates for load serving entities to establish a rate for every customer within the state, to no mandatory rules, as can be seen in Figure \ref{fig:nem}. The purchase rate in NEM programs are typically fixed retail rates, while NEB programs may have varying prices based on the underlying wholesale electricity price. In both cases, the compensation is adjusted to estimate the utility's avoided cost, based on the offset by net energy exports from DERs onto the broader grid. % over a specified time interval, most commonly monthly meter readings and billings
In some states such as Texas, retail net metering is not allowed, so the purchase rate is set to be a lower ``wholesale'' rate, which compensates the electricity import without subsidizing its production costs \citep{xcelTexas}.

While net metering has gained considerable traction in US markets, NEM policies overcompensate DGs, often providing a premium of up to 2-3 times the energy value. The purchase rate is typically comparable to the retail rate, despite the fact that the energy generation component only makes up half to a third of the rate, and NEM customers rely on the grid 24/7 for backup and do not contribute to grid maintenance \citep{NEM2020,NEMelephant}. The price for distribution system management is then offloaded onto non-NEM customers, and supports rate increases by utilities, furthering the social imbalance in electricity prices. The purchase rate can be even higher than the retail rate in states promoting solar PV uptake. In this way, NEM policies subsidize private solar, at the expense of non-solar owners, resulting in a reverse Robin Hood effect \citep{NEMForbes,NEMCGO}. These inequalities have spurred some regulatory reform efforts \citep{NEMelephant}, including the use of NEB in some states such as Texas, charging NEM customers a fee for grid maintenance, lowering the purchase rate, or pushing for compensation at the wholesale rate. Regardless of the purchase rate however, none of these compensation schemes fully incorporate locational or time price differentials, virtue of averaged tariff rates. These stagnant rates with limited volatility limit the adoption of dispatchable resources, such as storage, across the distribution system. Further, treating these behind-the-meter resources as primarily load modification resources limits the services they can provide to the grid, particularly in grid stability provisions \citep{pecanStreet}.

\subsection{Participation in Wholesale Markets}
In addition to retail programs and compensation schemes, participation of DERs in wholesale markets has been at the forefront of regulatory activity, and is further bolstered by FERC Order 2222 which requires ISO/RTOs to create DER aggregators as a market participant class. In this section we aim to cover a few diverse participation models. % and refer the reader to additional resources \textcolor{red}{[cite]} for further information. 

\subsubsection{Participation of DR Resources} \label{sec:wholesaleDR}
The participation of DR resources in wholesale markets began in the early 2000s, with more ISO/RTOs creating such programs to comply with FERC Orders 719 and 745. Most notably, CAISO has had aggregation programs for DR since 2001 \citep{GundlachCAISOReview}, and have since expanded the program to include more participation models, in the Proxy Demand Resource (PDR). In the PDR model, resources can bid load curtailment into both the energy and ancillary markets, with a minimum 100kW and 500kW load curtailment required respectively. The PDR model also includes the Reliability DR Resource (RDRR), in which curtailment is triggered only under emergency conditions. As of 2019, the Load Shift Resource (LSR) was introduced to allow bidirectional dispatch and reward resources for increasing consumption during negative pricing \citep{CAISOComparisonMatrix}. Many behind-the-meter resources participate through DR aggregations. However, DR are still classified as behind-the-meter load modification, not generation supply solutions, thus imposing a no-export rule on generators and storage participating in DR aggregations \citep{CAISOtoFERC}.

Participation of DR in the energy and reserves market was also introduced by ISO-NE in 2018, through the creation of the `active' DR class. Termed Price Responsive Demand, retail customers with DR capabilities are known as DR assets (DRAs), and are classified as either passive or active resources. Passive DRAs are nondispatchable, resources including energy efficiency resources and behind-the-meter solar PV, and cannot participate in energy or reserve markets. Active DRAs are dispatchable resources including load reduction, on-site generators, and storage. Similar to the CAISO model, these resources bid load reductions into the energy and reserve markets, and are cleared as resources comparable to generators. Aggregation is also permitted for active DRAs smaller than 5MW. Both active and passive resource classes can participate in the capacity market \citep{ISONEtoFERCder,ISONEtoFERCYoshimura,ISONE_DR}.

In PJM, DER participation in the wholesale market occurs through Curtailment Service Providers (CSP) which provide both emergency and energy resources. Emergency DR participate largely in the capacity market under the Reliability Pricing Model (RPM), with remuneration based on capacity commitments to be called on during emergency conditions, while energy DR can participate in both energy and ancillary markets. Energy DR are called upon to displace generators in the energy market when the wholesale price exceeds PJM net benefits price \citep{PJMconsumerfactsheet}.

Similar to the other participation models, MISO has allowed Aggregators of Retail Customers (ARCs) to bid into the wholesale market as reduction in demand since 2012. Based on the resource class, these ARCs can take part in the energy market, operating and planning reserves markets, and emergency response, and require minimum sizes of either 1MW for participation in energy and reserve markets, or 0.1MW for participation in emergency response. Behind-the-meter generation are subsumed within load modification resources, and cannot participate in energy and reserve markets, but can be used to meet resource adequacy requirements \citep{MISO_DER}.

\subsubsection{Participation of Small Generators and Storage}
CAISO has been at the forefront of DER integration by creating of the DER Provider (DERP) participation model. Established in 2015, this model allows aggregations to enable small-scale DERs, each $<1$MW in size, to collectively meet the minimum 0.5MW requirement to participate in the CAISO energy and ancillary markets. Aggregations are a new type of market resource, similar to a generating facility, and can bid into the market to be cleared as a single unit. Such aggregations can be composed of different resource types, and do not have to be geographically co-located. Rather, aggregations can span multiple transmission node connections, and therefore multiple pricing nodes, but must remain within electrically defined zones which have minimal price difference between the nodes, called sub-Load Aggregation Points (subLAP). Aggregations spanning multiple nodes cannot exceed 20MW. Each of the underlying resources are remunerated through a weighted average LMP across the pricing nodes of the aggregated resource, to reflect congestion related benefits from each resource \citep{CAISOtoFERC,CAISOparticipationGuideDER}. The DERP model has greatly influenced FERC's decision in issuing Order 2222. %\textcolor{blue}{CAISO's NGR creation for storage}

The participation model for ISO-NE does not use aggregators, but rather waives the minimum size requirement. The Settlement Only Resources (SOR) class consists of generators connected to the distribution system, and are less than 5MW. These resources participate in the RTM as price takers - they do not bid supply offers into the DAM or RTM; rather they self-dispatch and are paid the RT LMP when they produce energy. The SOR class can also participate in the capacity market if they are a minimum of 100kW \citep{ISONEtoFERCder,ISONEtoFERCYoshimura}. Resources participating in the capacity market are permitted to submit composite bids, in which resources with seasonal capacities can be aggregated to meet the year-round availability requirement. This allows summer-only distributed generation to couple with winter-only resources, widening the participation model for seasonal DERs \citep{ISONE_capacityresources}. In 2019, ISO-NE became the first to permit hybrid resources to participate in capacity market auctions, with the 2022-2023 auction clearing a bid for an aggregated residential solar-plus-storage resource, using a virtual power plant (VPP) model \citep{ISONEcapacitysolar}. The clearing price for this 14th auction was the lowest in the auction's history, at \$2 compared to \$7.03 in 2016 and \$3.80 in 2019 \citep{ISONEcapacityauction2}, continuing the downward trend in capacity prices, driven primarily by increased participation of solar resources supported by batteries \citep{ISONEcapacityauction1,ISONEcapacityauction3}.
%\textcolor{blue}{[Make some comment on PJM's proposed changes for participation? MISO doesn't have any non-DR models but is looking for ways to adjust for storage, need to look into NYISO]}

% \subsubsection{notes}
% DAM and RTM
% - ISO-NE SORs paid at RT LMP, no bidding
% - ISO-NE active DR (controllable loads, dispatchable BTM gen/storage) bid demand-reduction supply offers into DAM and RTM 

% Ancillary services
% - ISO-NE active DR (controllable loads, dispatchable BTM gen/storage) can participate in operating reserves

% Capacity market
% - ISO-NE SOR with minimum 100kW (up to 5MW)
% - ISO-NE passive DR and active DR

% (CAISO, MISO, ISONE, NYISO)

\subsubsection{Enrollment into Wholesale Programs}
Despite the existence of these models, there is limited participation of DERs in wholesale markets. In a review of the CAISO DERP aggregation model conducted in 2018, CAISO only had four participants registered in the DERP program, of which none had begun participating in either energy or ancillary markets. Interviewed active and potential participants indicated that participation in the wholesale markets would likely be limited to short- or medium-term, due to limited profitability. Further, DERs such as behind-the-meter storage are not well supported in the DERP model, which requires 24/7 settlement, prohibiting resources from stepping out when electricity prices are too high and discouraging DERs that were acquired primarily to meet on-site energy needs \citep{GundlachCAISOReview}. In the ISO-NE region, only 40\% of solar PV resources were participating in the wholesale market in 2019 \citep{ISONEtoFERCder}, though retail compensation schemes including NEM policies are thought to have contributed to the rapid growth of distributed solar \citep{GundlachCAISOReview}. While PJM has a large capacity of DER participation in wholesale markets compared to other regions, an estimated 7GW of DER potential still weren't participating in 2019. The share of DERs participating in PJM DR programs has also been decreasing since 2017, and DER participation as a DR resource in the energy markets has been decreasing since 2014. Further, of the locations within PJM with behind-the-meter resources like generation and batteries, most do not have export access; less than 5\% of these resources participate in either retail or wholesale activities, of which less than 8\% participate in wholesale markets \citep{PJMDERreport2019}. Similarly in MISO, 43\% of unregistered DERs are solar PV \citep{MISO_DER2020report}.

% \subsection{Other projects/participation models}
% \textcolor{blue}{other proposed models and pilot projects on transactive energy}

\subsection{Summary of Barriers for DERs for Market Participation}
A summary of the inefficiencies and/or barriers to participation of the above programs follows.
% Each of the programs discussed in this section introduce inefficiencies and/or barriers to participation. These are summarized below:
\begin{itemize}
    \item \textbf{Static pricing:} Temporal and locational pricing is not available to realize the flexible and unique nature of grid services from DERs. The limited variability of pricing signals in retail programs such as NEM and DR performance payments limits the adoption of resources which can respond quickly and dynamically to local conditions and provide grid-level support.
    \item \textbf{Voluntary enrollment:} Effective demand response programs must incentivize behavioral changes throughout the day, not just during performance periods. Essential to this is the increased participation in TOU rates, which are primarily opt-in programs, whose success enrollment success depends on promotion by utility companies. Although enrollment into TVP programs has been increasing since 2013, only a small fraction of retail customers are enrolled. With an estimated 200 GW of flexible load by 2030, widespread adoption of TVP programs by retail customers, especially EV owners, is necessary in realizing this potential. \citep{FERC_AMI2019} 
    \item \textbf{Competing retail and wholesale programs:} Current market designs do not permit participation in both retail and wholesale programs. As a result, these programs compete with one another for DER enrollment. For example, the DERP program inadvertently competes with the wholesale PDR and retail NEM, of which the latter two programs are less costly to participate in \citep{GundlachCAISOReview}. However, both of these structures provide limited services to the grid, subject to no-export rules, limited ancillary market participation, and introduce barriers to the entry of storage. The misalignment of the different tariff structures limits the profitability for DER services \citep{SunSpecDERReport}.
    \item \textbf{Prohibitive technical requirements:} The technical requirements from DERs participating in both retail and wholesale programs are limiting in two ways: misalignment with grid services, and economic barriers. For example, Rule 21 interconnection standards require residential DERs to have only hourly or day-ahead functionalities, thus limiting their usefulness, particularly in their ability to provide stability provisions. Further, the metering and telemetry requirement for aggregated resources is the same as for traditional generators, despite their different capacities and capabilities. Resource aggregations also typically do not benefit from economies of scale. In CAISO's DERP program, each DER which is part of the aggregation must install its own revenue meter, which introduces prohibitive costs for small operators \citep{GundlachCAISOReview}.
    \item \textbf{Prohibitive regulatory requirements:} Interconnection rules and procedures vary between retail and wholesale level participation, which creates a barrier of entry for DERs already participating at the retail level to enter the wholesale market, despite the creation of aggregation programs. For example, under CAISO, if a resource is connected by Rule 21 (for which rules vary between utility distribution companies) and wants to now participate in wholesale, it must reapply under WDAT, which is structured for conventional generators and often allocates the cost of technologies to the resource. Second, there is no standardized communication protocol across retail and wholesale spaces, limiting DER participation in aggregations or under VPP models. Third, wholesale programs which allow DERs to provide services to the grid are limiting, in that they often require all-year and 24/7 participation. Under such models, resources cannot step out of the market when desired\footnote{This rule limits arbitrage opportunities, which aligns with ISO/RTOs need for reliable and transparent market participation.}, unless they enroll in DR programs. This is limiting for DERs, especially those which are behind-the-meter and serve on-site load: a storage device discharging locally (i.e. not injecting power to the grid) when the LMP is negative must make payments to the wholesale market \citep{GundlachCAISOReview}.
\end{itemize}

\section{Proposed Retail Market Design} \label{sec:market}
As DER penetration continues to increase, better coordination of these resources is warranted. There is a need for increased temporal and locational granularity in electricity pricing, innovative ancillary products, an expansion of market derivatives to include more grid services, and an alignment of retail and wholesale markets through coordinated tariff structures and market clearing schemes, and are the emerging responsibilities for DSOs \citep{IRENA2019briefDSO}. In our market design, the DSO is responsible for overseeing the participation of DERs in a retail market, through which DERs are scheduled and remunerated at real-time prices. The market is composed of a (1) real-time energy market which schedules DERs and determine market settlements; and (2) an ancillary services market which balances load across primary feeders. In this paper, we limit our focus to the energy market.
% The DSO carries out the overall scheduling through two markets: (1) a retail energy market that consist of DER scheduling and real-time market settlements; and (2) an ancillary market that has oversight over transactions for alert conditions, between primary feeders in the distribution network. In this paper, we will focus on the energy market.
% , proposing a market structure composed of an energy market and ancillary market, built on the foundations of a market mechanism in which the temporal and locational capabilities of DERs can be priced.
% \textcolor{blue}{In Section \ref{sec:policy}, we then extend the discussion to how this structure lays the groundwork for the other DSO tasks, in particular coordination with the TSO and the expansion of market derivatives. (or do some of this in Section 4.4?)}

\par The DSO is composed of two entities, the Workers (DSO-W) and Representatives (DSO-R), which reside at the substation and primary feeder respectively. The DSO-Rs oversee the energy market and aggregate data of the DERs under their purview; and the DSO-Ws operate the ancillary market and aggregate information from the DSO-Rs. While the DSO acts as a data aggregator, it does not bid into the WEM on behalf of its DERs like an aggregation company or transmission level resource. Rather, the DSO can be viewed as a proactive utility in the sense that it accepts the Locational Marginal Price (LMP) as traditionally determined by the WEM, and optimally makes use of the DERs within the distribution network to maximize economic efficiency and other network-level objectives. In doing so, the DSO requests service from the WEM only for net loads beyond the DER capabilities, and compensates/charges the DERs for their services/usage at the d-LMP.
% The DSO is composed of two entities, the Workers (DSO-W) and Representatives (DSO-R). The DSO-Rs carry out the energy market, and aggregate data of the DERs under their purview. The DSO-Ws are responsible for the operation of the ancillary services market, with support from the DSO-Rs. While the DSO acts as an aggregator of data, it does not bid into the WEM on behalf of its DERs like an aggregation company or transmission level resource. Rather, the DSO should be thought of as a proactive utility in the sense that it accepts the Locational Marginal Price (LMP) as traditionally determined by the WEM, and optimally makes use of the DERs within the distribution network to maximize economic efficiency and other network-level objectives. In doing so, the DSO requests service from the WEM only for net loads beyond the DER capabilities, and compensates/charges the DERs for their services/usage at the d-LMP.

\begin{figure}
	\centering
	\includegraphics[trim={0.65cm 0cm 0cm 0cm},clip,scale=0.2]{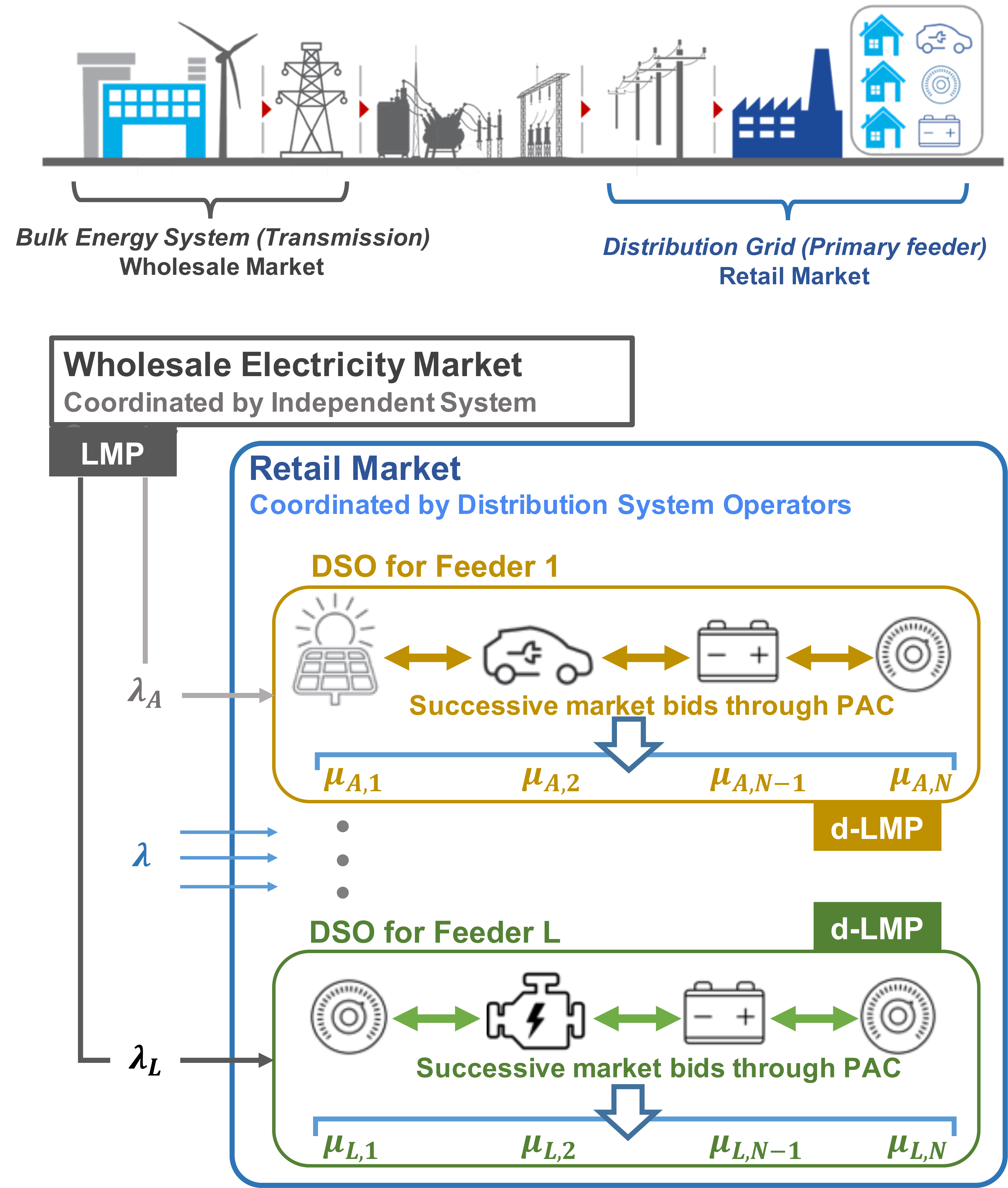}
	\caption{Proposed retail market structure}
	\label{fig:DGmarketFull}
\end{figure}

A schematic of the operation of the proposed retail market is shown in Fig. \ref{fig:DGmarketFull}. 

% A schematic of both the physical layer and the market layer of the proposed retail market is shown in Fig. \ref{fig:DGmarketFull}. All data communication coming from the WEM is shown with solid arrows, while data communication moving upstream through the DSO is shown with dashed arrows.

% \begin{figure*}
% 	\centering
% 	\includegraphics[trim={0.3cm 3.2cm 0.3cm 3.2cm},clip,scale=0.55]{figs/marketdiagram_whitepaper.pdf}
% 	\caption{Proposed retail market structure}
% 	\label{fig:DGmarketFull}
% \end{figure*} 

% trim={0.3cm 3.2cm 0.3cm 3.2cm},clip,

\subsection{Operation of Real-Time Energy Market}
The energy market is a highly distributed local real-time market carried out by the DSO-Rs. The market operates at the primary feeder level (4 to 35 kV level); any DERs and uncontrollable loads at the secondary feeder level and below are represented through aggregators, through which they can participate in this retail market. Each primary feeder has its own DSO-R which oversees the energy market. To simplify the discussion, we model every bus in the physical network layer as an independent agent participating in the market layer, which represents all the DERs located at and/or below that node\footnote{This is not a technical limitation of the proposed structure. Multiple neighbouring nodes can choose to be represented by the same agent, which would then have access to all required operational data and pricing information. A detailed discussion of such an agent is beyond the scope of this paper.}. These DERs include both behind-the-meter resources and those connected directly to the distribution grid, including DR, DGs, and storage. 

Each agent is equipped with the necessary computational and communication infrastructure to participate in the market, which is built upon a distributed optimization algorithm called PAC (for technical details see \cite{haiderRM,RomvaryTAC,RomvaryThesis}). Using this algorithm, each agent self-schedules to minimize its expenses (equivalently maximize profit) while subjected to network constraints such as voltage limits, thermal line limits, and other DSO-level objectives, which are modeled through a non-linear convex optimal power flow formulation. The DER dispatch schedules and d-LMPs are determined by repeated negotiations between neighbouring agents using peer-to-peer communication, which are carried out autonomously using PAC. During every negotiation, the PAC algorithm requires each agent to communicate its proposed load or generation setpoint and variables pertaining to the network's physical constraints (such as voltage and current) to its neighbours. These interactions will determine the d-LMP (retail cost of electricity per kW) of each agent, based on marginal cost arguments. After reaching an agreement with its neighbours, each agent enters into a bi-lateral agreement with its DSO-R, committing to deliver or consume the decided amount of power, at the d-LMP. The net load consumed by an agent will be charged at this d-LMP, and equivalently, the net generation by an agent will be remunerated at the d-LMP. Payments will be made to/from the DSO-R. The above transactions proceed in a parallel fashion across all agents reporting to a DSO-R. It is recommended that this retail energy market has a shorter clearing period than the WEM clearing. As non-dispatchable resources displace conventional generation, market clearing times must be faster and more flexible, in order to better reflect the highly temporal nature of renewable resources and to accommodate updated forecasts \citep{NASEM_report,PoplavskayaBalancingMarkets}. For example, the WEM in ISO-NE clears every 5 minutes, so the proposed retail market could clear every minute. The market clearing time however, can be freely selected to suit the needs of each ISO/RTO. It should be noted here that our proposed retail market includes a market for both real and reactive power. Although reactive power markets don't currently exist, even in the WEM - due to issues of price volatility and market power concerns - the increase in DER penetration together with enabling technologies such as smart inverters has the potential to realize efficient reactive power market designs.

\subsection{Validation of Proposed Market Design} \label{sec:validation}
We benchmark the proposed market operation against four operating models, wherein the utility purchases power from the WEM at the wholesale price and sells to customers at a fixed retail price, as is currently done in the US. The first of these is a `Traditional' model where there is no DER utilization. The `No Export' model realizes DR as continuous DER operation, rather than only during specific call windows which are a limitation of the programs introduced in Sec.~\ref{sec:DRprograms}, and retain the no-export rule for behind-the-meter resources from wholesale market participation models (see Sec.~\ref{sec:wholesaleDR}). Retail compensation schemes NEM and NEB discussed in Sec.~\ref{sec:retailcomp} are used in the `Retail\_M' and `Retail\_B' models. \newline 
\textbf{Traditional:} There is no DER utilization within the network. All load is serviced by the utility. \newline
\textbf{No Export:} DG resources are used to offset local load and cannot export excess generation to the grid (excess generation is curtailed). All load from customers without DGs and any excess load of DG owners is serviced by the utility. \newline
\textbf{Retail\_M:} DG resources are used to offset local load and can export excess generation to the grid. Compensation for DGs follows NEM, at a fixed retail purchase rate. Any excess network load is serviced by the utility. \newline
\textbf{Retail\_B:} DG resources are used to offset local load and can export excess generation to the grid. Compensation for DGs follows NEB, at a fixed retail purchase rate. Any excess network load is serviced by the utility. 

% We benchmark the proposed market operation against a system where the utility purchases power from the WEM at the wholesale price and sells to customers at a retail price. In this model, we assume that DERs are not compensated for their services, as they are small-scale behind-the-meter resources. The metrics can be calculated as follows.

We use several metrics to validate the market performance using different stakeholder perspectives. They include the revenue for a DER owner, the cost for a customer consuming electricity, and the net revenue for the DSO. These metrics are calculated as follows. Real and reactive power are denoted as $P$ and $Q$, with superscripts G and L for generation and consumption respectively. Subscript $j$ denotes the j-th agent participating in the market. The wholesale LMP is denoted as $\lambda^\text{P}$, retail electricity prices as $\mu^\text{P}_\text{retail}$, retail purchase rate as $\mu^\text{P}_\text{retail-G}$, and the d-LMP for an agent $j$ as $\mu^\text{P}_j$ and $\mu^\text{Q}_j$. The baseline load for an agent $j$ is denoted as $P^\text{L0}_j$ and $Q^\text{L0}_j$. With this notation, we define the following quantities.\newline

\noindent \textit{Payment made to WEM, for purchasing power:} \newline
$\mathcal{C}_{\text{WEM}} = \lambda^{\text{P}}\sum_{j} P_j^{\text{L}}$ \label{eq:cost:WEM} \newline
\textit{Revenue earned from loads without proposed market:} \newline
$\mathcal{R}_{\text{load}}^\text{base}=\sum_{j}  \mu_\text{retail}^{\text{P}}P_j^{\text{L}}$ \label{eq:rev:loadnoDSO} \newline
\textit{Revenue earned from loads with proposed market:} \newline
$\mathcal{R}_{\text{load}}^\text{market}=\sum_{j}  \Big( \mu_j^{\text{P}}P_j^{\text{L}}+\mu_j^{\text{Q}}Q_j^{\text{L}}\Big)$ \label{eq:rev:loadDSO} \newline
\textit{Remuneration to distributed generators:} \newline
$\mathcal{R}_{\text{gen}}=\sum_{j}  \Big( \mu^{\text{P}}P_j^{\text{G}}+\mu^{\text{Q}}Q_j^{\text{G}}\Big)$ \label{eq:rem:dg} \newline 
where for traditional and no export cases $\mu^{\text{P}}=0$ and $\mu^{\text{Q}}=0$, for Retail\_M and Retail\_B cases $\mu^{\text{P}}=\mu_\text{retail-G}^{\text{P}}$ and $\mu^{\text{Q}}=0$, and for the proposed market $\mu^{\text{P}}=\mu_j^{\text{P}}$ and $\mu^{\text{Q}}=\mu_j^{\text{Q}}$. \newline
\noindent \textit{Remuneration to flexible loads:}\newline
$\mathcal{R}_{\text{flex}}=\sum_{j}  \Big(\mu_j^{\text{P}} \big(P_j^{\text{L0}}-P_j^{\text{L}}\big)+\mu_j^{\text{Q}} \big(Q_j^{\text{L0}}-Q_j^{\text{L}}\big)\Big)$ \label{eq:rem:flex} \newline

\noindent The metrics are then defined as: \newline 
\textbf{Revenue for DER owner:} 
$\mathcal{R}_\text{flex}$ and $\mathcal{R}_\text{gen}$ \newline
\textbf{Cost for consumer:}  $\mathcal{R}_{\text{load}}^\text{x}$ \newline
\textbf{Net revenue:} 
$\mathcal{P} = \mathcal{R}_{\text{load}}^\text{x} - \mathcal{R}_{\text{flex}} - \mathcal{R}_{\text{gen}} - \mathcal{C}_{\text{WEM}}$ \newline

% $\mathcal{S}_{j}=(\lambda_{\text{retail}}^{\text{P}}-\mu_j^{\text{P}})P_j^{\text{L}} -\mu_j^{\text{Q}}Q_j^{\text{L}}$ \label{eq:savings:cust} \newline 

% $\mathcal{P}_\text{DSO-increase} = \mathcal{P}_\text{DSO}-\mathcal{P}_\text{no-DSO} \\
% = \big[\mathcal{R}_{\text{load}}^\text{market} - \mathcal{R}_{\text{flex}} - \mathcal{R}_{\text{gen}} - \mathcal{C}_{\text{WEM}}\big]-\big[\mathcal{R}_{\text{load}} - \mathcal{C}_{\text{WEM}}\big]$

% eversource NEM August 2020 list: https://www.eversource.com/content/docs/default-source/builders-contractors/nm-credits-boston.pdf?sfvrsn=c162c262_40

% \textcolor{blue}{Note: two different cases for payment made to WEM - consider net metering as here where demand response is not remunerated but people still adjust load anyways, and then where people do not adjust load. }

In the numerical exercise that follows, the market operation has been simulated over a 24 hour period, on the IEEE-123 node network, which is a primary distribution feeder model. The network data was modified to be a balanced 3-phase distribution network, and DERs were added to the network \citep{haiderRM}. All loads are assumed to be capable of DR (in real power). About 10\% of the nodes in the grid are assumed to have local generating capabilities, with almost 70\% of the total network load capable of being met by the total nameplate generation. Market reports from ISO-NE operations provide the five-minute approved LMPs (wholesale price) \citep{isoneLMP}, and five-minute total recorded electricity demand from which the time-dependent demand ratio $\alpha(t)$ is calculated, for August 25, 2020 \citep{isoneD}. Load data from the IEEE datasheet provides the upper bound on load forecast, with the real-time forecast varying as per $\alpha(t)$. Retail data from Eversource in Massachusetts is used for the benchmark scenarios, with $\mu_\text{retail}^P = \$0.114$/kWh (generation service charge for basic service), and $\mu_\text{retail-G}^P = \$0.192$/kWh (Class I solar/wind under Residential R-1 tariff) \citep{eversourceNEM}. 
% \textcolor{blue}{figures}

\begin{figure}
	\subfloat[d-LMP Real Power. \label{fig:muQ}]{\includegraphics[trim={1cm 3cm 2cm 5.2cm},clip,scale=0.3]{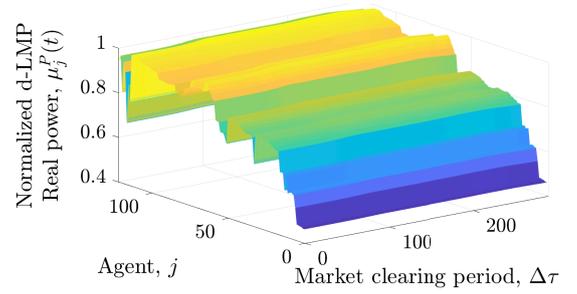}}
	\\
	\subfloat[d-LMP Reactive power\label{fig:muQ}]{\includegraphics[trim={0.5cm 3cm 2cm 4.8cm},clip,scale=0.3]{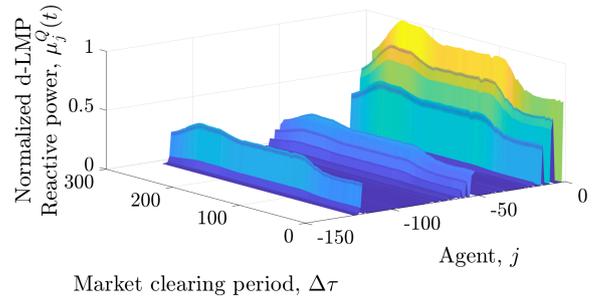}}
	\caption{Locational-temporal variation in retail price, using PAC algorithm and proposed market}
	\label{fig:muResults}
\end{figure}

\begin{figure}
    \centering
    \includegraphics[trim={1.5cm 4.2cm 0.5cm 4.3cm},clip,scale=0.32]{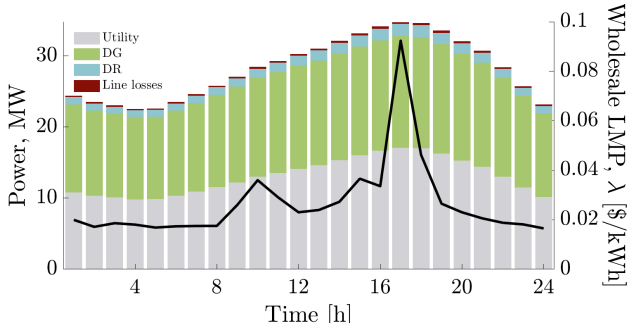}
    \caption{Schedule for resources using proposed RM and PAC. The forecasted load is serviced by the utility, DGs, and DR curtailment. Additional power is purchased due to losses in the network.}
    \label{fig:kWBreakdownRM}
\end{figure}

\begin{figure}
    \centering
    \includegraphics[trim={1.9cm 1cm 3.3cm 1.5cm},clip,scale=0.28]{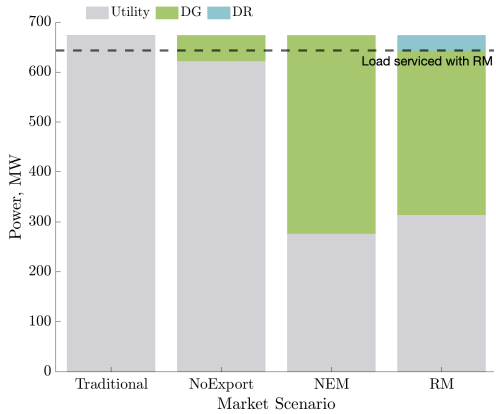}
    \caption{Resources utilized across different market scenarios. The forecasted load is serviced by the utility, DGs, and DR curtailment.}
    \label{fig:kWBreakdownAll}
\end{figure}

\begin{figure*}
    \centering
    \includegraphics[trim={1.75cm 3.75cm 1cm 3.75cm},clip,scale=0.6]{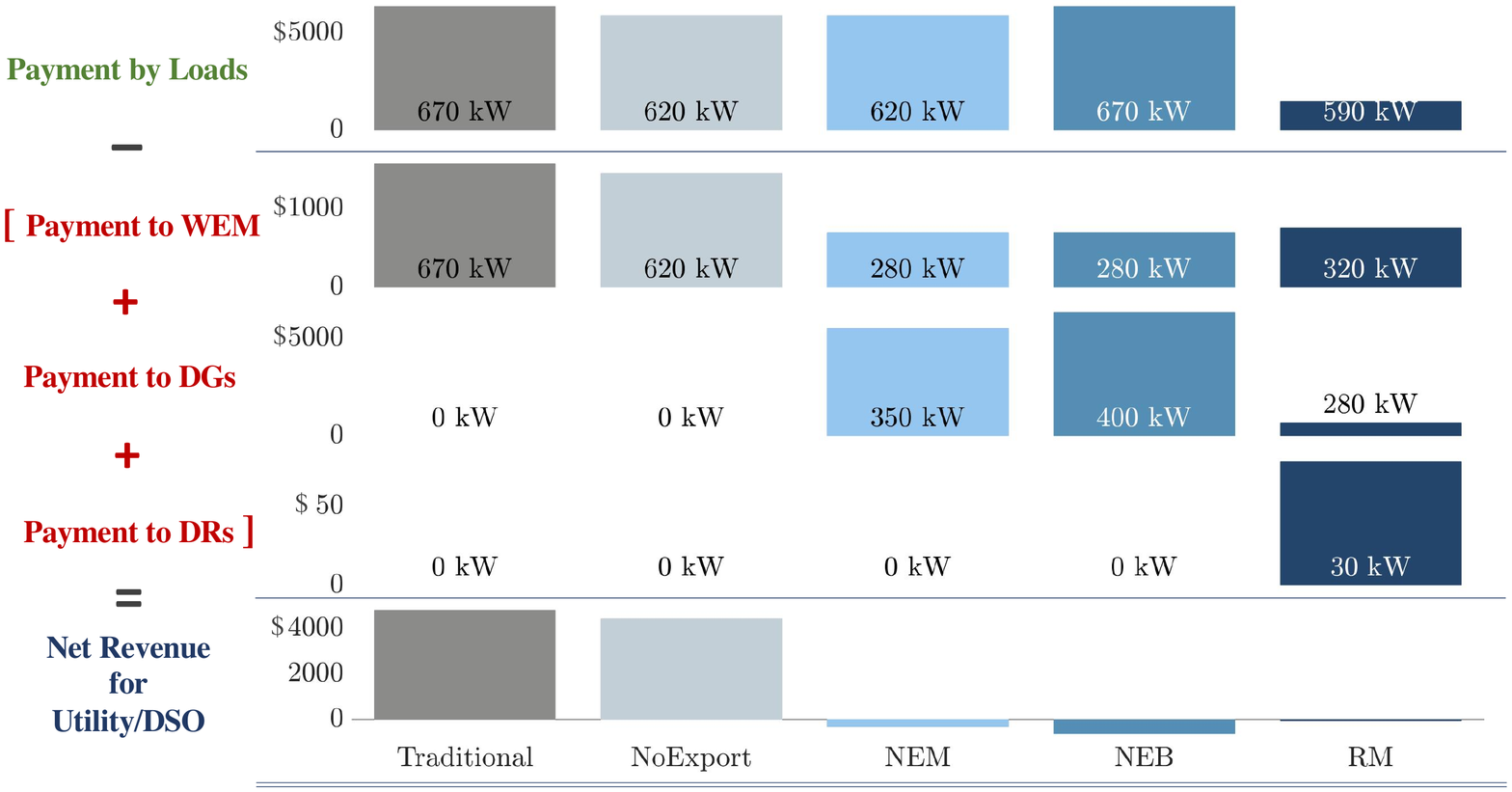}
    \caption{A breakdown of the net revenue calculation under all scenarios, with all revenue and cost in USD. The kW value used to calculate the amount is annotated on the bars - these are the load serviced by the utility, power purchased from the WEM, load serviced by the DGs, and load curtailment by DRs.}
    \label{fig:profitBreakdown}
\end{figure*}

Results from the simulation are presented in Figures~\ref{fig:muResults}-\ref{fig:profitBreakdown}. The normalized retail prices from the proposed energy market, which are calculated using PAC, are shown in Fig.~\ref{fig:muResults}. There is a high locational variation in both prices, with $\mu_j^P$ prices varying by a factor of 2 within the same period. The results also show temporal variation for prices at the same node, with higher variation in $\mu_j^Q$ which sees a factor of 3.8 between highest and lowest prices throughout the day. The temporal variation of $\mu_j^Q$ roughly follows the demand ratio $\alpha(t)$, with higher prices during higher load periods. This is likely because the DG units were configured to provide only real power and all reactive power load must be met by the utility purchasing power from the WEM; to stabilize the prices, all DGs must have reactive power capabilities, such as solar PV with smart inverters. This is in line with the revised Rule 21 interconnection procedures in CAISO, which require DG units to be equipped with smart inverters prior to their approval. In comparison, there is lower temporal variability in $\mu_j^P$; this is likely due to the modeling choices (DGs with continuous output and a fixed percentage of curtailable load at all times of the day) and the smooth demand curve. More realistic data including the variability in renewable generation such as a day with passing cloud cover, networks with high loading conditions, and more granular modeling of DR capabilities may increase the volatility of the real-time price. The retail market allows the DSO to price these spatial-temporal variations and realize the true value of energy services provided by DERs. It may not be desirable to expose customers to these volatile prices, which can be remedied by more traditional TVP techniques, which average prices over a period such as an hour. 

The aggregated hourly schedule determined by the energy market is shown in Fig.~\ref{fig:kWBreakdownRM}. The forecasted network load is serviced by the utility purchasing power from the WEM (in grey), DGs serving both onsite load and exporting power to the grid (in green), and curtailment from demand response (in blue). The graph also shows the total power loss due to electrical resistance (in burgundy). The wholesale price $\lambda$ is also plotted (black line).
The maximum LMP coincides with peak network load in hour 17, during which both DR and DG utilization is at a maximum. Periods of low demand and low wholesale prices have lower resource utilization, as purchasing power from the WEM is comparable to remunerating a DER at $\mu_j^P$, with an average LMP of \$0.0267/kWh and average $\mu_j^P$ of \$0.0291/kWh. The aggregated resource utilization for each market operation benchmark and the proposed retail market is shown in Fig.\ref{fig:kWBreakdownAll}. The dashed line shows the total load serviced under the proposed market operation, which is lower than the benchmark cases which do not have DR enabled. Both the Traditional and No Export scenarios fail to utilize DERs, and while the Retail\_M scenario does use DGs, there is no coordination of resources to achieve economic and energy efficiency. 

A detailed comparison of the cost of market operation is shown in Fig.~\ref{fig:profitBreakdown}. Both the Traditional and No Export scenarios result in large profits for the utility, due to the large difference between the retail and wholesale prices of electricity. Both Retail\_M and Retail\_B result in a loss for the utility. While these retail compensation structures are currently used in US electricity markets, the high retail purchase rate means the utility is not only overcompensating the DGs, but that under high penetration of these DG resources, this participation model becomes uneconomical. One option is to provide lower purchase rates, however deciding the value of the energy service being provided is challenging. Another option is to enable participation at the wholesale level, but this continues to be challenging for small resources, even through aggregator models. Most notably in Fig.~\ref{fig:profitBreakdown}, all the quantities for the retail market scenario are significantly lower than of the Traditional and No Export case, and only comparable to the Retail\_M/Retail\_B cases for the cost of electricity from the WEM. Despite serving the same load, the proposed retail market is able to do this at a much lower retail cost: an average of 0.0291 \$/kWh, compared to the current utility retail price of 0.114 \$/kWh. Rather than simply making a large profit, the proposed DSO is building social equity and redistributing wealth through socialization of the profit. With the retail market, the true value of energy is recovered, which results in lower electricity costs for consumers and lower compensation for DERs, while ensuring power balance and economic efficiency in the market. 

\subsection{Enabling Technologies} \label{sec:tech}
To implement the proposed retail market and support the secure bidding between agents, communication infrastructure in the distribution grid is needed. 
%This can be classified into (a) the physical network infrastructure to ensure connectivity; and (b) the protocols to support market function - including communication and data coordination. %These are discussed below.
The existing communication infrastructure in the bulk energy grid, which are used to support operations and wholesale market functions, consists of optical fibers connecting control centers to substations, and multiple protocols like Synchronous Optical Networking (SONET) which enable fast and secure grid communications. At the sub-transmission and distribution level, Wide Area Networks (WAN) or LTE based networks are used in conjunction with various protocols such as DNP3 (legacy) or IEC 61850, to measure and control the grid. Market functions, such as submitting bids, are typically performed over the Internet, using various authentication mechanisms to ensure security. %This communication infrastructure enables wholesale market operations by providing the physical network and standardized operating procedures.

To implement the retail market and realize the successive autonomous bidding mechanism of the PAC algorithm for each agent, the existing communication infrastructure can be fully utilized. Currently, SCADA networks provide limited communication and control in distribution systems; however, this is rapidly changing with the proliferation of AMI systems. While AMIs were initially deployed to aid in grid operations, their use has evolved to support market functions such as TVP~\citep{FERC_AMI2019}. Further, with the advent of the Internet of Things (IoT), connectivity and communication with grid edge resources and loads is enabled without the need to build additional infrastructure, providing more visibility and control capabilities. Various communication protocols such as ZigBee, Modbus, IEEE Std 802.15.4, and PLC standards allow for communication with AMI devices. Recently, the IEEE 2030.5 Smart Energy Profile (SEP) 2.0 standard, which provides a framework for monitoring and control of DER assets, has been gaining traction with grid operators, and has been suggested as the standardized communication protocol for DER aggregation programs. For example, CAISO outlines in their Common Smart Inverter Profile (CSIP) how SEP should be implemented to meet Rule 21, requiring DERs to have monitoring and reporting capabilities, and grid support functionalities such as Volt-VAr Control (VVC) \citep{SunSpecDERReport}. With these technologies, a PAC-based retail market can be realized by leveraging the grid-edge intelligence and connectivity of resources.

\section{Conclusion and Policy Implications} \label{sec:policy}
A retail market operating within the distribution grid enables the participation of small scale resources, which include, among others, DG, DR, and storage including EVs. These DERs can be compensated for the services they provide at a real-time rate with such a market, based on their marginal cost of operation and current grid conditions. Such an approach also allows resources to operate more dynamically and eliminates the no-export rules for behind-the-meter DERs\footnote{where technically feasible, i.e. grid constraints are satisfied}, so they can generate, reduce load, or even increase load as needed by the network. As DER penetration continues to increase, technology costs reduce, and subsidies for these resources are removed, new incentives for DER participation in markets is required. This can be achieved through new revenue streams from retail markets. In this paper we proposed a retail market structure using a distributed optimization algorithm capable of solving for the optimal dispatch and d-LMPs, while leveraging grid-edge intelligence and peer-to-peer communication \citep{haiderRM,RomvaryTAC}. 
% , such as net metering and feed-in tariffs for solar PV, will be removed, 

The proposed energy market can also be augmented with an ancillary market \citep{haiderRM}, by allowing DSO-W to coordinate DSO-Rs to ensure balance of supply and demand under service disruptions commitments. In doing so, the flexibility of the DERs under the purview of the DSO-R are able to participate in balancing and are compensated for their fast reacting capabilities. 
These distributed resources with computational abilities can then be utilized for grid resiliency in the face of large outage events, extreme weather, and cyber attacks. Designing these derivatives and understanding the operational overlap between energy, ancillary, and `voltage' services is necessary to fully realize smart coordination of DERs and leverage their flexibility. 
%Regulatory oversight is needed to determine the types of grid support functions that can be performed by DERs, and adapted based on the resource location. 

The hierarchical structure also lends itself to localized energy markets within the secondary feeder, realized through technologies such as blockchain-enabled peer-to-peer energy trading. %\}{[cite examples?]}.
At this level, submetering can be used for load disaggregation, particularly for co-located resources such as EVs and loads. This notion of hierarchical markets and operations requires coordination between interfacing markets: the local and retail market at the interface of the primary feeder, and the retail and wholesale market at the interface of the transmission substation. Coordination can be realized by communicating net load/generation and forecasts, sharing operating status, or even bidding mechanisms. In this latter case, at the DSO-TSO interface the DSO acts as an aggregator, bidding the net load/generation for the network into the WEM. Co-optimization can be carried out through iterative schemes, to determine the optimal schedule of bulk resources and DSO aggregators, and determine the LMP at the DSO node. In this way, the DSO is no longer a price taker but an active market player capable of setting wholesale prices, and the DERs can better respond to bulk energy system changes. 

From a regulatory perspective, an aggregator participation model for DERs into the wholesale market needs to be further analyzed, especially with FERC Order 2222: should DERs be able to participate only in the retail market, or should direct wholesale participation be allowed, and if so, can resources participate in \textit{both} markets simultaneously? If able to participate in either market, can resources freely move between them, or will they be required to provide 24/7 service to a single market for a duration of time, say a year? Is a reactive power market with contained volatility realizable? In each of these participation models, resource adequacy and market fairness come into question: are participation models centered around the flexibility of DERs (as required by FERC Order 841) fair to traditional bulk resources which cannot readily step in and out of markets, and are being pushed out of the energy market by lower cost renewables - and, in future, DERs - but are still needed on standby to provide fast ramping? Alignment is also needed with balancing and reserve markets to integrate DERs alongside a retail market, else market gaming between intraday and balancing markets will persist \citep{JustGermanyBalancing}. Our proposal for an allocation of tasks among the wholesale and retail markets is this: The need for new or updated capacity models and compensation for standby generators must be addressed at the wholesale level. The task of DER integration and compensation must be addressed at the retail level. There needs to be appropriate coordination between these two markets. Our proposed retail market is a first step in answering all of the above questions.

% In this way, the introduction of a retail market creates a natural separation in grid services, allowing the ISO/RTOs to focus on system-wide resource adequacy questions, and leaves the task of DER integration and compensation to the DSO at the local level. Furthermore, coordination protocols are needed between the transmission and distribution networks to address the communication gap between technical and market perspectives introduced by FERC Order 2222 for aggregator participation in markets without corresponding power flow calculations, and for procurement markets and planning activities. The introduction of a retail market is a necessary step for DER integration, as it develops the necessary infrastructure and visibility into the distribution grid needed for such coordination.

% While the physical infrastructure and technology to support retail markets exists, additional regulatory policy is needed to standardize procedures and communication protocols, especially between wholesale and retail spaces. Market participation rules and standardized protocols will govern the landscape under which DERs are integrated into the grid.

A few statements need to be made regarding the lower costs for customers that can be realized using the proposed retail market. This is in sharp contrast to the current regulatory structures in which retail prices are increasing despite the drop in levelized cost for renewables and decreasing -- and sometimes even negative -- wholesale prices \citep{ForbesElectricityPrices}. Efficient pricing and resource coordination at the distribution level permit these lower costs. As we look towards real-time pricing, however, we must also consider rate equity across different socioeconomic classes, to ensure fair access of electricity \citep{BurgerBook}. Regulator concerns about exposing customers to the price volatility of real-time rates can be alleviated by ensuring tariff designs are equitable, and by employing hybrid TVP models such as Block-and-Index pricing to enable risk hedging. The lower retail price also results in a lower revenue stream for the DSO. However, we note that this is not endemic to the proposed retail market, but rather a reality of the modern electricity grid with negative electricity prices (frequently occurring in US states of California and Texas, and Germany), high renewable curtailment, and unprecedented ramp rates as in the famous California `duck curve'. This is already manifesting in systems with high renewable penetration at the transmission level: the share prices for the three largest utilities in Germany have dropped by 45\% to 66\% between 2010 and 2016, and other utilities in Western Europe have similarly lost market value over the past decade \citep{MITEI_UoF}. While this may seem concerning for the future of the utility as we know it, it is a reality stemming from the misalignment of incentives present in the current utility business model. The utility model and corresponding rate structures must be redesigned to shift away from a commoditized market with capital expenditures and energy sales as the main revenue stream, and towards  performance-based ratemaking (PBR) where utility revenue is instead based on achieving performance metrics and other non-investment factors. Although the compensation mechanism for resources, in particular generators, will likely become increasingly complex, this new business model can help realign revenue with state RPS and energy goals, by supporting utility investment into NWA and more efficient grid utilization. Retail electricity prices then better reflect both the quality of service for customers, and the performance and responsiveness of utilities to government mandates \citep{NRELPBR,ForbesPBR}. The extent to which the composition of revenue stream will change and the resulting decrease in retail electricity prices depends heavily on the geographical location. Factors such as renewable penetration at the transmission level, DER penetration, interdependence of electricity prices and commodity prices (such as natural gas in the Northeast US), climate, wholesale market structure and capacity procurement, and regional RPS goals and policies can result in different performance metrics for a single retail market design. Of equal importance is analyzing how regulatory and policy changes can impact the business model structure for providing electricity services \citep{BurgerBusinessModels}, and ensuring high enrollment of DERs to increase market liquidity \citep{WeberLiquidity}.

Finally, the design of the retail market must uphold and support both state and federal policy objectives. The optimal market design requires consideration of both short- and long-term incentives for all market participants \citep{WeberLiquidity}. More analysis is needed to determine how the proposed market structure can promote investment into energy efficiency, grid reliability, and clean energy. Assessing the impact of carbon pricing and environmental costs, and accounting for externalities such as air quality and healthcare costs is also necessary \citep{BellGillReview}. The interaction between the electricity and natural gas markets also needs to be better understood, especially in areas where gas is used for both bulk electricity production and home heating, as in the Northeast US. Another interesting concept is the realization of a `thermal market' whereby DR is also enabled for thermal loads, such as space and water heating, which traditionally rely on gas. As we look to electrify more sectors of the economy, including heating, transportation, and manufacturing, the interaction of these networks must be accounted for. A smart city approach can better integrate electricity consumption, EVs, and thermal loads, to achieve higher operational efficiency and lower costs.

\printcredits

\bibliographystyle{cas-model2-names}

\bibliography{cas-refs}

\begin{thebibliography}{70}
\expandafter\ifx\csname natexlab\endcsname\relax\def\natexlab#1{#1}\fi
\providecommand{\url}[1]{\texttt{#1}}
\providecommand{\href}[2]{#2}
\providecommand{\path}[1]{#1}
\providecommand{\DOIprefix}{doi:}
\providecommand{\ArXivprefix}{arXiv:}
\providecommand{\URLprefix}{URL: }
\providecommand{\Pubmedprefix}{pmid:}
\providecommand{\doi}[1]{\href{http://dx.doi.org/#1}{\path{#1}}}
\providecommand{\Pubmed}[1]{\href{pmid:#1}{\path{#1}}}
\providecommand{\bibinfo}[2]{#2}
\ifx\xfnm\relax \def\xfnm[#1]{\unskip,\space#1}\fi
%Type = Article
\bibitem[{Aggarwal(2018)}]{ForbesPBR}
\bibinfo{author}{Aggarwal, S.}, \bibinfo{year}{2018}.
\newblock \bibinfo{title}{America's utility of the future forms around
  performance-based regulation}.
\newblock \bibinfo{journal}{Forbes} .
%Type = Techreport
\bibitem[{Anisie et~al.(2019)Anisie, Boshell and Ocenic}]{IRENA2019briefDSO}
\bibinfo{author}{Anisie, A.}, \bibinfo{author}{Boshell, F.},
  \bibinfo{author}{Ocenic, E.}, \bibinfo{year}{2019}.
\newblock \bibinfo{title}{Future Role of Distribution System Operators:
  Innovation Landscape Brief}.
\newblock \bibinfo{type}{Technical Report}. International Renewable Energy
  Agency.
%Type = Article
\bibitem[{Bell and Gill(2018)}]{BellGillReview}
\bibinfo{author}{Bell, K.}, \bibinfo{author}{Gill, S.}, \bibinfo{year}{2018}.
\newblock \bibinfo{title}{Delivering a highly distributed electricity system:
  Technical, regulatory and policy challenges}.
\newblock \bibinfo{journal}{Energy policy} \bibinfo{volume}{113}.
%Type = Article
\bibitem[{Borenstein(2002)}]{borenstein2002trouble}
\bibinfo{author}{Borenstein, S.}, \bibinfo{year}{2002}.
\newblock \bibinfo{title}{The trouble with electricity markets: understanding
  {C}alifornia's restructuring disaster}.
\newblock \bibinfo{journal}{Journal of economic perspectives}
  \bibinfo{volume}{16}, \bibinfo{pages}{191--211}.
%Type = Article
\bibitem[{Borenstein(2014)}]{haasDR2014}
\bibinfo{author}{Borenstein, S.}, \bibinfo{year}{2014}.
\newblock \bibinfo{title}{Money for nothing?}
\newblock \bibinfo{journal}{Energy Institute at {HAAS}} .
%Type = Book
\bibitem[{B{\"o}s(2015)}]{bos2015pricing}
\bibinfo{author}{B{\"o}s, D.}, \bibinfo{year}{2015}.
\newblock \bibinfo{title}{Pricing and price regulation: an economic theory for
  public enterprises and public utilities}. volume~\bibinfo{volume}{34}.
\newblock \bibinfo{publisher}{Elsevier}.
%Type = Inbook
\bibitem[{Burger et~al.(2019)Burger, Schneider, Botterud and
  Pérez-Arriaga}]{BurgerBook}
\bibinfo{author}{Burger, S.}, \bibinfo{author}{Schneider, I.},
  \bibinfo{author}{Botterud, A.}, \bibinfo{author}{Pérez-Arriaga, I.},
  \bibinfo{year}{2019}.
\newblock \bibinfo{title}{Fair, Equitable, and Efficient Tariffs in the
  Presence of Distributed Energy Resources}.
\newblock pp. \bibinfo{pages}{155--188}.
\newblock \DOIprefix\doi{10.1016/B978-0-12-816835-6.00008-5}.
%Type = Article
\bibitem[{Burger and Luke(2017)}]{BurgerBusinessModels}
\bibinfo{author}{Burger, S.P.}, \bibinfo{author}{Luke, M.},
  \bibinfo{year}{2017}.
\newblock \bibinfo{title}{Business models for distributed energy resources: A
  review and empirical analysis}.
\newblock \bibinfo{journal}{Energy policy} \bibinfo{volume}{109}.
%Type = Misc
\bibitem[{CAISO()}]{CAISOComparisonMatrix}
\bibinfo{author}{CAISO}, .
\newblock \bibinfo{title}{Summary comparison matrix}.
\newblock
  \bibinfo{howpublished}{\url{http://www.caiso.com/Documents/ParticipationComparison-ProxyDemand-DistributedEnergy-Storage.pdf}}.
%Type = Misc
\bibitem[{CAISO(2016)}]{CAISOparticipationGuideDER}
\bibinfo{author}{CAISO}, \bibinfo{year}{2016}.
\newblock \bibinfo{title}{Distributed energy resource provider participation
  guide with checklist}.
\newblock
  \bibinfo{howpublished}{\url{http://www.caiso.com/Documents/DistributedEnergyResourceProviderParticipationGuideandChecklist.pdf}}.
%Type = Misc
\bibitem[{{E}nergy~{S}olutions {C}enter()}]{NEM_NEB}
\bibinfo{author}{{E}nergy~{S}olutions {C}enter, C.}, .
\newblock \bibinfo{title}{Net metering and net billing}.
\newblock
  \bibinfo{howpublished}{\url{https://cleanenergysolutions.org/instruments/net-metering-net-billing\#:~:text=Net\%20metering\%20credits\%20can\%20be,purchased\%20during\%20other\%20time\%20periods.&text=The\%20primary\%20difference\%20between\%20net,the\%20grid\%20under\%20net\%20billing.}}
%Type = Misc
\bibitem[{Chernyakhovskiy et~al.(2016)Chernyakhovskiy, Tian, McLaren, Miller
  and Geller}]{NRELRegreport}
\bibinfo{author}{Chernyakhovskiy, I.}, \bibinfo{author}{Tian, T.},
  \bibinfo{author}{McLaren, J.}, \bibinfo{author}{Miller, M.},
  \bibinfo{author}{Geller, N.}, \bibinfo{year}{2016}.
\newblock \bibinfo{title}{{U.S.} laws and regulations for renewable energy grid
  interconnections}.
%Type = Misc
\bibitem[{{P}aso~{E}lectric {C}ompany(2020)}]{elpaso_DR}
\bibinfo{author}{{P}aso~{E}lectric {C}ompany, E.}, \bibinfo{year}{2020}.
\newblock \bibinfo{title}{El {P}aso {E}lectric {C}ompany’s 2020 load
  management program}.
\newblock
  \bibinfo{howpublished}{\url{https://www.epelectric.com/files/html/Energy_efficiency/Energy_Efficiency_Program_Manuals/2020_Program_Manuals/2020\%20TX\%20Load\%20Management\%20Program\%20Manual.pdf}}.
%Type = Misc
\bibitem[{CPower(2020)}]{DR_PJM_cpower}
\bibinfo{author}{CPower}, \bibinfo{year}{2020}.
\newblock \bibinfo{title}{Understanding {PJM} capacity demand response
  changes}.
\newblock
  \bibinfo{howpublished}{\url{https://cpowerenergymanagement.com/help/pjm-dr-changes/\#:~:text=PJM\%20and\%20its\%20Emergency\%20Capacity,Capacity\%20Performance\%20(CP)\%20programs.}}
%Type = Techreport
\bibitem[{DSIRE(2019)}]{dsireNEM2019}
\bibinfo{author}{DSIRE}, \bibinfo{year}{2019}.
\newblock \bibinfo{title}{Net Metering Policies by State}.
\newblock \bibinfo{type}{Technical Report}. Database of State Incentives for
  Renewables and Efficiency.
%Type = Article
\bibitem[{Eckhouse and Martin(2019)}]{ISONEcapacityauction3}
\bibinfo{author}{Eckhouse, B.}, \bibinfo{author}{Martin, C.},
  \bibinfo{year}{2019}.
\newblock \bibinfo{title}{Residential solar and storage to participate in {N}ew
  {E}ngland wholesale energy capacity market}.
\newblock \bibinfo{journal}{Bloomberg News Editors} .
%Type = Misc
\bibitem[{EERE()}]{all_TVP_DR}
\bibinfo{author}{EERE, U.}, .
\newblock \bibinfo{title}{Demand response and time-variable pricing programs}.
\newblock
  \bibinfo{howpublished}{\url{https://www.energy.gov/eere/femp/demand-response-and-time-variable-pricing-programs}}.
%Type = Misc
\bibitem[{EIA(2011)}]{ISO_LMP}
\bibinfo{author}{EIA}, \bibinfo{year}{2011}.
\newblock \bibinfo{title}{Wholesale power price maps reflect real-time
  constraints on transmission of electricity}.
\newblock
  \bibinfo{howpublished}{\url{https://www.eia.gov/todayinenergy/detail.php?id=3150\#}}.
%Type = Misc
\bibitem[{Energy(2016)}]{DR_NHPUC}
\bibinfo{author}{Energy, E.}, \bibinfo{year}{2016}.
\newblock \bibinfo{title}{{NHPUC} no. 9 – electricity delivery; public
  service company of {N}ew {H}ampshire {DBA} {E}versource {E}nergy}.
\newblock
  \bibinfo{howpublished}{\url{https://www.eversource.com/content/docs/default-source/rates-tariffs/electric-delivery-service-tariff-nh.pdf?sfvrsn=7fb7f062_52}}.
%Type = Misc
\bibitem[{Energy(2020)}]{eversourceNEM}
\bibinfo{author}{Energy, E.}, \bibinfo{year}{2020}.
\newblock \bibinfo{title}{Eastern {M}assachusetts net metering credit pricing}.
\newblock
  \bibinfo{howpublished}{\url{https://www.eversource.com/content/docs/default-source/builders-contractors/nm-credits-boston.pdf?sfvrsn=c162c262_40}}.
%Type = Misc
\bibitem[{of~Energy Efficiency \& Renewable~Energy()}]{coned_TVP}
\bibinfo{author}{of~Energy Efficiency \& Renewable~Energy, O.}, .
\newblock \bibinfo{title}{Time-of-use rates}.
\newblock
  \bibinfo{howpublished}{\url{https://www.coned.com/en/save-money/energy-saving-programs/time-of-use}}.
%Type = Incollection
\bibitem[{Faruqui(2012)}]{faruqui2012ethics}
\bibinfo{author}{Faruqui, A.}, \bibinfo{year}{2012}.
\newblock \bibinfo{title}{The ethics of dynamic pricing}, in:
  \bibinfo{booktitle}{Smart Grid}. \bibinfo{publisher}{Elsevier}, pp.
  \bibinfo{pages}{61--83}.
%Type = Techreport
\bibitem[{FERC(2018)}]{fercDRReport2018}
\bibinfo{author}{FERC}, \bibinfo{year}{2018}.
\newblock \bibinfo{title}{2018 Assessment of Demand Response and Advanced
  Metering}.
\newblock \bibinfo{type}{Technical Report}. Federal Energy Regulatory
  Commission.
%Type = Misc
\bibitem[{{FERC}(2020)}]{ferc2222}
\bibinfo{author}{{FERC}}, \bibinfo{year}{2020}.
\newblock \bibinfo{title}{Participation of {D}istributed {E}nergy {R}esource
  {A}ggregations in markets operated by {R}egional {T}ransmission
  {O}rganizations and {I}ndependent {S}ystem {O}perators}.
%Type = Misc
\bibitem[{Foley et~al.(2019)Foley, Blomberg and Kakley}]{ISONEcapacityauction1}
\bibinfo{author}{Foley, E.}, \bibinfo{author}{Blomberg, M.},
  \bibinfo{author}{Kakley, M.}, \bibinfo{year}{2019}.
\newblock \bibinfo{title}{{N}ew {E}ngland's forward capacity auction closes
  with adequate power system resources for 2022-2023}.
\newblock
  \bibinfo{howpublished}{\url{https://www.iso-ne.com/static-assets/documents/2019/02/20190206_pr_fca13_initial_results.pdf}}.
%Type = Misc
\bibitem[{Foster et~al.(2019)Foster, Bialecki, Burns, Kathan, Lee and
  Peirovi}]{FERC_AMI2019}
\bibinfo{author}{Foster, B.}, \bibinfo{author}{Bialecki, T.},
  \bibinfo{author}{Burns, D.}, \bibinfo{author}{Kathan, D.},
  \bibinfo{author}{Lee, M.P.}, \bibinfo{author}{Peirovi, S.},
  \bibinfo{year}{2019}.
\newblock \bibinfo{title}{2019 assessment of demand response and advanced
  metering}.
\newblock
  \bibinfo{howpublished}{\url{https://www.ferc.gov/sites/default/files/2020-04/DR-AM-Report2019_2.pdf}}.
%Type = Article
\bibitem[{Gerard et~al.(2018)Gerard, Puente and Six}]{gerard2018coordination}
\bibinfo{author}{Gerard, H.}, \bibinfo{author}{Puente, E.I.R.},
  \bibinfo{author}{Six, D.}, \bibinfo{year}{2018}.
\newblock \bibinfo{title}{Coordination between transmission and distribution
  system operators in the electricity sector: A conceptual framework}.
\newblock \bibinfo{journal}{Utilities Policy} \bibinfo{volume}{50},
  \bibinfo{pages}{40--48}.
%Type = Article
\bibitem[{Gheorghiu(2019)}]{ISONEcapacitysolar}
\bibinfo{author}{Gheorghiu, I.}, \bibinfo{year}{2019}.
\newblock \bibinfo{title}{Residential solar+storage breaks new ground as
  {S}unrun wins {ISO-NE} capacity contract}.
\newblock \bibinfo{journal}{UtilityDive} .
%Type = Article
\bibitem[{Gundlach and Webb(2018)}]{GundlachCAISOReview}
\bibinfo{author}{Gundlach, J.}, \bibinfo{author}{Webb, R.},
  \bibinfo{year}{2018}.
\newblock \bibinfo{title}{Distributed energy resource participation in
  wholesale markets: Lessons from the {C}alifornia {ISO}}.
\newblock \bibinfo{journal}{Energy Law Journal}
  \DOIprefix\doi{10.7916/D8CR79T5}.
%Type = Article
\bibitem[{Haider et~al.(2020)Haider, Baros, Wasa, Romvary, Uchida and
  Annaswamy}]{haiderRM}
\bibinfo{author}{Haider, R.}, \bibinfo{author}{Baros, S.},
  \bibinfo{author}{Wasa, Y.}, \bibinfo{author}{Romvary, J.},
  \bibinfo{author}{Uchida, K.}, \bibinfo{author}{Annaswamy, A.M.},
  \bibinfo{year}{2020}.
\newblock \bibinfo{title}{Towards a retail market for distribution grids}.
\newblock \bibinfo{journal}{IEEE Transactions on Smart Grid}
  \DOIprefix\doi{10.1109/TSG.2020.2996565}.
%Type = Misc
\bibitem[{Hinson(2019)}]{pecanStreet}
\bibinfo{author}{Hinson, S.}, \bibinfo{year}{2019}.
\newblock \bibinfo{title}{What {C}alifornia's {R}ule 21 gets right and wrong
  for residential solar}.
\newblock
  \bibinfo{howpublished}{\url{https://www.pecanstreet.org/2019/04/what-californias-rule-21-gets-right-and-wrong-for-residential-solar/}}.
%Type = Article
\bibitem[{Hogan(2010)}]{hogan2010fairness}
\bibinfo{author}{Hogan, W.W.}, \bibinfo{year}{2010}.
\newblock \bibinfo{title}{Fairness and dynamic pricing: comments}.
\newblock \bibinfo{journal}{The Electricity Journal} \bibinfo{volume}{23},
  \bibinfo{pages}{28--35}.
%Type = Techreport
\bibitem[{{IRENA}(2019)}]{IRENA2019innovation}
\bibinfo{author}{{IRENA}}, \bibinfo{year}{2019}.
\newblock \bibinfo{title}{Innovation Landscape for a Renewable-powered future:
  Solutions to integrate variable renewables}.
\newblock \bibinfo{type}{Technical Report}. International Renewable Energy
  Agency.
%Type = Misc
\bibitem[{ISO-NE(2019a)}]{ISONE_DR}
\bibinfo{author}{ISO-NE}, \bibinfo{year}{2019}a.
\newblock \bibinfo{title}{About demand resources}.
\newblock
  \bibinfo{howpublished}{\url{https://www.iso-ne.com/markets-operations/markets/demand-resources/about}}.
%Type = Misc
\bibitem[{ISO-NE(2019b)}]{ISONEtoFERCder}
\bibinfo{author}{ISO-NE}, \bibinfo{year}{2019}b.
\newblock \bibinfo{title}{Response to letter dated {S}eptember 5,2019 regarding
  participation of distributed energy resource aggregations in markets operated
  by regional transmission organizations and independent system operators}.
\newblock
  \bibinfo{howpublished}{\url{https://www.iso-ne.com/static-assets/documents/2019/10/rm18-9_resp_to_der_data_req.pdf}}.
%Type = Misc
\bibitem[{ISO-NE(2020a)}]{isoneD}
\bibinfo{author}{ISO-NE}, \bibinfo{year}{2020}a.
\newblock \bibinfo{title}{Energy, load, and demand reports}.
\newblock
  \bibinfo{howpublished}{\url{https://www.iso-ne.com/isoexpress/web/reports/load-and-demand/-/tree/dmnd-five-minute-sys}}.
%Type = Misc
\bibitem[{ISO-NE(2020b)}]{isoneLMP}
\bibinfo{author}{ISO-NE}, \bibinfo{year}{2020}b.
\newblock \bibinfo{title}{Pricing reports}.
\newblock
  \bibinfo{howpublished}{\url{https://www.iso-ne.com/isoexpress/web/reports/pricing/-/tree/lmps-rt-five-minute-final}}.
%Type = Misc
\bibitem[{ISO-NE(2020c)}]{ISONEcapacityauction2}
\bibinfo{author}{ISO-NE}, \bibinfo{year}{2020}c.
\newblock \bibinfo{title}{Results of the annual forward capacity auctions}.
\newblock
  \bibinfo{howpublished}{\url{https://www.iso-ne.com/about/key-stats/markets\#fcaresults}}.
%Type = Article
\bibitem[{John()}]{GTMferc841}
\bibinfo{author}{John, J.S.}, .
\newblock \bibinfo{title}{`{E}normous step' for energy storage as court upholds
  {FERC} {O}rder 841, opening wholesale markets}.
\newblock \bibinfo{journal}{GreenTechMedia} \URLprefix
  \url{https://www.greentechmedia.com/articles/read/court-upholds-ferc-order-841-opening-wholesale-markets-to-energy-storage}.
%Type = Article
\bibitem[{Just and Weber(2015)}]{JustGermanyBalancing}
\bibinfo{author}{Just, S.}, \bibinfo{author}{Weber, C.}, \bibinfo{year}{2015}.
\newblock \bibinfo{title}{Strategic behavior in the {G}erman balancing energy
  mechanism: incentives, evidence, costs and solutions}.
\newblock \bibinfo{journal}{Journal of Regulatory Economics}
  \bibinfo{volume}{48}.
%Type = Techreport
\bibitem[{Littel et~al.(2017)Littel, Kadoch, Baker, Bharvirkar, Dupuy,
  Hausauer, Linvill, Migden-Ostrander, Rosenow, Xuan, Zinaman and
  Logan}]{NRELPBR}
\bibinfo{author}{Littel, D.}, \bibinfo{author}{Kadoch, C.},
  \bibinfo{author}{Baker, P.}, \bibinfo{author}{Bharvirkar, R.},
  \bibinfo{author}{Dupuy, M.}, \bibinfo{author}{Hausauer, B.},
  \bibinfo{author}{Linvill, C.}, \bibinfo{author}{Migden-Ostrander, J.},
  \bibinfo{author}{Rosenow, J.}, \bibinfo{author}{Xuan, W.},
  \bibinfo{author}{Zinaman, O.}, \bibinfo{author}{Logan, J.},
  \bibinfo{year}{2017}.
\newblock \bibinfo{title}{Next-Generation Performance-Based Regulation}.
\newblock \bibinfo{type}{Technical Report}. NREL.
%Type = Misc
\bibitem[{MISO(2020a)}]{MISO_DER}
\bibinfo{author}{MISO}, \bibinfo{year}{2020}a.
\newblock \bibinfo{title}{Der markets workshop}.
\newblock
  \bibinfo{howpublished}{\url{https://cdn.misoenergy.org/20200331\%20DER\%20Workshop\%20MISO\%20Summary\%20Presentation440314.pdf}}.
%Type = Misc
\bibitem[{MISO(2020b)}]{MISO_DER2020report}
\bibinfo{author}{MISO}, \bibinfo{year}{2020}b.
\newblock \bibinfo{title}{{MISO} and {DER}: Framing and discussion document}.
\newblock
  \bibinfo{howpublished}{\url{https://cdn.misoenergy.org/DER\%20Framing\%20Report\%202019397951.pdf}}.
%Type = Techreport
\bibitem[{MITEI(2016)}]{MITEI_UoF}
\bibinfo{author}{MITEI}, \bibinfo{year}{2016}.
\newblock \bibinfo{title}{Utility of the Future}.
\newblock \bibinfo{type}{Technical Report}. MIT Energy Initiative.
%Type = Article
\bibitem[{Murray(2019)}]{ForbesElectricityPrices}
\bibinfo{author}{Murray, B.}, \bibinfo{year}{2019}.
\newblock \bibinfo{title}{The paradox of declining renewable costs and rising
  electricity prices}.
\newblock \bibinfo{journal}{Forbes} .
%Type = Book
\bibitem[{NASEM(2021)}]{NASEM_report}
\bibinfo{author}{NASEM}, \bibinfo{year}{2021}.
\newblock \bibinfo{title}{Legal and Regulatory Issues that shape the Electric
  System}.
\newblock \bibinfo{publisher}{The {N}ational {A}cademies {P}ress}.
%Type = Misc
\bibitem[{Nichols and Lehman(2019)}]{ISONE_capacityresources}
\bibinfo{author}{Nichols, J.}, \bibinfo{author}{Lehman, S.},
  \bibinfo{year}{2019}.
\newblock \bibinfo{title}{Having a capacity supply obligation lesson 2c:
  Introduction to capacity resources}.
\newblock
  \bibinfo{howpublished}{\url{https://www.iso-ne.com/static-assets/documents/2019/10/20191021-fcm101-lesson-2C-intro-capacity-resources_PRINT.pdf}}.
%Type = Misc
\bibitem[{NIST(2017)}]{NISTtransactive}
\bibinfo{author}{NIST}, \bibinfo{year}{2017}.
\newblock \bibinfo{title}{Transactive energy: An overview}.
\newblock
  \bibinfo{howpublished}{\url{https://www.nist.gov/engineering-laboratory/smart-grid/hot-topics/transactive-energy-overview}}.
%Type = Inbook
\bibitem[{Nudell et~al.(2019)Nudell, Annaswamy, Lian, Kalsi and
  D'Achiardi}]{Nudell2019}
\bibinfo{author}{Nudell, T.R.}, \bibinfo{author}{Annaswamy, A.M.},
  \bibinfo{author}{Lian, J.}, \bibinfo{author}{Kalsi, K.},
  \bibinfo{author}{D'Achiardi, D.}, \bibinfo{year}{2019}.
\newblock \bibinfo{title}{Electricity Markets in the United States: A Brief
  History, Current Operations, and Trends}. \bibinfo{publisher}{Springer
  International Publishing}, \bibinfo{address}{Cham}.
\newblock pp. \bibinfo{pages}{3--27}.
\newblock \DOIprefix\doi{10.1007/978-3-319-98310-3_1}.
%Type = Article
\bibitem[{Pepper(2013)}]{baltimoreBaseball}
\bibinfo{author}{Pepper, T.}, \bibinfo{year}{2013}.
\newblock \bibinfo{title}{{FERC} settles investigation concerning demand
  response products in {PJM}}.
\newblock \bibinfo{journal}{Washington Energy Report} .
%Type = Misc
\bibitem[{PJM()}]{PJMconsumerfactsheet}
\bibinfo{author}{PJM}, .
\newblock \bibinfo{title}{Retail electricity consumer opportunities for demand
  response in pjm's wholesale markets}.
\newblock
  \bibinfo{howpublished}{\url{https://www.pjm.com/~/media/markets-ops/dsr/end-use-customer-fact-sheet.ashx}}.
%Type = Misc
\bibitem[{PJM(2020)}]{PJMDERreport2019}
\bibinfo{author}{PJM}, \bibinfo{year}{2020}.
\newblock \bibinfo{title}{2019 distributed energy resources ({DER}) that
  participate in {PJM} markets as demand response}.
\newblock
  \bibinfo{howpublished}{\url{https://www.pjm.com/-/media/markets-ops/demand-response/2019-der-annual-report.ashx?la=en}}.
%Type = Article
\bibitem[{Poplavskaya and Vries(2019)}]{PoplavskayaBalancingMarkets}
\bibinfo{author}{Poplavskaya, K.}, \bibinfo{author}{Vries, L.D.},
  \bibinfo{year}{2019}.
\newblock \bibinfo{title}{Distributed energy resources and the organized
  balancing market: A symbiosis yet? case of three european balancing markets}.
\newblock \bibinfo{journal}{Energy policy} \bibinfo{volume}{126}.
%Type = Article
\bibitem[{Puckett(2020)}]{NEM2020}
\bibinfo{author}{Puckett, J.}, \bibinfo{year}{2020}.
\newblock \bibinfo{title}{Time to make net metering a net positive}.
\newblock \bibinfo{journal}{Real Clear Energy} .
%Type = Article
\bibitem[{Ritchie(2016)}]{NEMForbes}
\bibinfo{author}{Ritchie, E.J.}, \bibinfo{year}{2016}.
\newblock \bibinfo{title}{The solar net metering controversy: Who pays for
  energy subsidies?}
\newblock \bibinfo{journal}{Forbes} .
%Type = Phdthesis
\bibitem[{Romvary(2018)}]{RomvaryThesis}
\bibinfo{author}{Romvary, J.}, \bibinfo{year}{2018}.
\newblock \bibinfo{title}{A proximal atomic coordination algorithm for
  distributed optimization in distribution grids}.
\newblock Ph.D. thesis. Massachusetts Institute of Technology.
%Type = Article
\bibitem[{Romvary et~al.(2020)Romvary, Ferro, Haider and
  Annaswamy}]{RomvaryTAC}
\bibinfo{author}{Romvary, J.}, \bibinfo{author}{Ferro, G.},
  \bibinfo{author}{Haider, R.}, \bibinfo{author}{Annaswamy, A.M.},
  \bibinfo{year}{2020}.
\newblock \bibinfo{title}{A distributed proximal atomic coordination
  algorithm}.
\newblock \bibinfo{journal}{IEEE Transactions on Automatic Control
  (provisionally accepted)} .
%Type = Article
\bibitem[{Ruester et~al.(2014)Ruester, Schwenen, Batlle and
  P{\'e}rez-Arriaga}]{ruester2014distribution}
\bibinfo{author}{Ruester, S.}, \bibinfo{author}{Schwenen, S.},
  \bibinfo{author}{Batlle, C.}, \bibinfo{author}{P{\'e}rez-Arriaga, I.},
  \bibinfo{year}{2014}.
\newblock \bibinfo{title}{From distribution networks to smart distribution
  systems: Rethinking the regulation of {E}uropean electricity {DSO}s}.
\newblock \bibinfo{journal}{Utilities Policy} \bibinfo{volume}{31},
  \bibinfo{pages}{229--237}.
%Type = Article
\bibitem[{Smith et~al.(2018)Smith, Patty and Colton}]{NEMCGO}
\bibinfo{author}{Smith, J.T.}, \bibinfo{author}{Patty, G.},
  \bibinfo{author}{Colton, K.}, \bibinfo{year}{2018}.
\newblock \bibinfo{title}{Net metering in the {S}tates: A primer on reforms to
  avoid regressive effects and encourage competition}.
\newblock \bibinfo{journal}{The Center for Growth and Opportunity} .
%Type = Techreport
\bibitem[{Tansy et~al.(2018)Tansy, Nelson, Moy and Martinez}]{SunSpecDERReport}
\bibinfo{author}{Tansy, T.}, \bibinfo{author}{Nelson, R.},
  \bibinfo{author}{Moy, K.}, \bibinfo{author}{Martinez, S.},
  \bibinfo{year}{2018}.
\newblock \bibinfo{title}{Analysis Report of Wholesale Energy Market
  Participation by Distributed Energy Resources ({DER}s) in {C}alifornia}.
\newblock \bibinfo{type}{Technical Report}. SunSpec Alliance.
%Type = Article
\bibitem[{Trabish(2020)}]{DRnews_CAISO}
\bibinfo{author}{Trabish, H.K.}, \bibinfo{year}{2020}.
\newblock \bibinfo{title}{Demand response failed {C}alifornia 20 years ago; the
  state's recent outages may have redeemed it}.
\newblock \bibinfo{journal}{UtilityDive} .
%Type = Misc
\bibitem[{Ulmer et~al.(2018)Ulmer, Collanton, Ivancovich and
  Weaver}]{CAISOtoFERC}
\bibinfo{author}{Ulmer, A.}, \bibinfo{author}{Collanton, R.E.},
  \bibinfo{author}{Ivancovich, A.}, \bibinfo{author}{Weaver, W.},
  \bibinfo{year}{2018}.
\newblock \bibinfo{title}{Post-technical conference comments of the
  {C}alifornia {I}ndependent {S}ystem {O}perator {C}orporation}.
%Type = Techreport
\bibitem[{{Vlerick Energy Centre}(2020)}]{KPMG-EUoutlook}
\bibinfo{author}{{Vlerick Energy Centre}}, \bibinfo{year}{2020}.
\newblock \bibinfo{title}{Outlook on the {E}uropean {DSO} Landscape 2020}.
\newblock \bibinfo{type}{Technical Report}. Vlerick Business School.
\newblock
  \bibinfo{note}{\url{https://home.kpmg/content/dam/kpmg/pdf/2016/05/Energy-Outlook-DSO-2020.pdf}}.
%Type = Article
\bibitem[{Weber(2010)}]{WeberLiquidity}
\bibinfo{author}{Weber, C.}, \bibinfo{year}{2010}.
\newblock \bibinfo{title}{Adequate intraday market design to enable the
  integration of wind energy into the european power systems}.
\newblock \bibinfo{journal}{Energy policy} \bibinfo{volume}{38}.
%Type = Article
\bibitem[{Wolak(2006)}]{anaheimCPP}
\bibinfo{author}{Wolak, F.A.}, \bibinfo{year}{2006}.
\newblock \bibinfo{title}{Residential customer response to real-time pricing:
  The {A}naheim {C}ritical-{P}eak {P}ricing experiment} .
%Type = Article
\bibitem[{Wolfram(2017)}]{haasDR2017}
\bibinfo{author}{Wolfram, C.}, \bibinfo{year}{2017}.
\newblock \bibinfo{title}{The problem with demand response}.
\newblock \bibinfo{journal}{Energy Institute at {HAAS}} .
%Type = Article
\bibitem[{Wood(2016)}]{NEMelephant}
\bibinfo{author}{Wood, L.V.}, \bibinfo{year}{2016}.
\newblock \bibinfo{title}{Why net energy metering results in a subsidy: The
  elephant in the room}.
\newblock \bibinfo{journal}{Brookings} .
%Type = Misc
\bibitem[{XcelEnergy()}]{xcelTexas}
\bibinfo{author}{XcelEnergy}, .
\newblock \bibinfo{title}{Texas solar and private generation}.
\newblock
  \bibinfo{howpublished}{\url{https://www.xcelenergy.com/staticfiles/xe-responsive/Programs\%20and\%20Rebates/Residential/Complete-information-diagram-application.pdf}}.
%Type = Misc
\bibitem[{Yoshimura(2018)}]{ISONEtoFERCYoshimura}
\bibinfo{author}{Yoshimura, H.}, \bibinfo{year}{2018}.
\newblock \bibinfo{title}{Statement of {H}enry {Y}oshimura, {ISO} {N}ew
  {E}ngland {I}nc.}
\newblock
  \bibinfo{howpublished}{\url{https://www.iso-ne.com/static-assets/documents/2018/04/rm18-9_4-9-18_yoshimura_statement.pdf}}.
%Type = Techreport
\bibitem[{Zinaman et~al.(2015)Zinaman, Miller, Adil, Arent, Aggarwal, Bipath,
  Linvill, David, Kauffman, Futch, Arcos, Valenzuela, Martinot, Bazilian and
  Pillai}]{NRELfuture}
\bibinfo{author}{Zinaman, O.}, \bibinfo{author}{Miller, M.},
  \bibinfo{author}{Adil, A.}, \bibinfo{author}{Arent, D.},
  \bibinfo{author}{Aggarwal, S.}, \bibinfo{author}{Bipath, M.},
  \bibinfo{author}{Linvill, C.}, \bibinfo{author}{David, A.},
  \bibinfo{author}{Kauffman, R.}, \bibinfo{author}{Futch, M.},
  \bibinfo{author}{Arcos, E.V.}, \bibinfo{author}{Valenzuela, J.M.},
  \bibinfo{author}{Martinot, E.}, \bibinfo{author}{Bazilian, M.},
  \bibinfo{author}{Pillai, R.K.}, \bibinfo{year}{2015}.
\newblock \bibinfo{title}{Power Systems of the Future: A 21st century power
  partnership thought leadership report}.
\newblock \bibinfo{type}{Technical Report}. NREL.

\end{thebibliography}

\end{document}